\documentclass[
aps,
prd,
twocolumn,preprintnumbers,
superscriptaddress,
nofootinbib,
floatfix,
]{revtex4-2}

\pdfoutput=1

\usepackage{graphicx}
\usepackage{amsmath,amsfonts,amssymb}
\usepackage{xcolor}
\usepackage[normalem]{ulem}
\usepackage{hyperref}
\usepackage{comment}

\hypersetup{
  colorlinks = true,
  citecolor  = red,
  linkcolor  = blue,
  urlcolor   = blue
}

\definecolor{ao(english)}{rgb}{0.0,0.5,0.0}

\newcommand{\noi}{\noindent}
\newcommand{\beq}{\begin{equation}}
\newcommand{\eeq}{\end{equation}}
\newcommand{\bea}{\begin{eqnarray}}
\newcommand{\eea}{\end{eqnarray}}

\newcommand{\Trh}{T_{\rm RH}}
\newcommand{\Tmax}{T_{\rm max}}
\newcommand{\Tosc}{T_{\rm osc}}
\newcommand{\arh}{a_{\rm RH}}
\newcommand{\aend}{a_{\rm end}}
\newcommand{\rhoend}{\rho_{\rm end}}
\newcommand{\Hend}{H_{\rm end}}

\graphicspath{{./figs/}}

\begin{document}

 \preprint{UMN--TH--4601/26, FTPI--MINN--26/18}

\title{Spectator Dark Matter and the Higgs Portal}

\author{Stephen E. Henrich}
\email{henri455@umn.edu}
\affiliation{
William I. Fine Theoretical Physics Institute,
School of Physics and Astronomy,
University of Minnesota,
Minneapolis, Minnesota 55455, USA
}

\author{Yann Mambrini}
\email{yann.mambrini@ijclab.in2p3.fr}
\affiliation{
Universit\'e Paris-Saclay,
CNRS/IN2P3,
IJCLab,
91405 Orsay,
France
}

\author{Keith Olive}
\email{olive@umn.edu}
\affiliation{
William I. Fine Theoretical Physics Institute,
School of Physics and Astronomy,
University of Minnesota,
Minneapolis, Minnesota 55455, USA
}

\begin{abstract}

Light scalar spectators generically acquire large stochastic fluctuations
during inflation, which can lead to a catastrophic overproduction of dark matter if the resulting primordial condensate survives until late times. We show that this inflationary memory can be sufficiently suppressed by an extremely feeble Higgs portal interaction with sub-freeze-in couplings ($10^{-28} \lesssim \lambda_{HS} \lesssim 10^{-9}$). The Higgs portal induces both a radiative self-interaction for the spectator as well as a thermal mass, both of which can lead to early onset of oscillations and efficient redshifting of the condensate's energy density without thermalizing the spectator or destroying the condensate. The Higgs portal also alters the stochastic equilibrium distribution of the spectator fluctuations, leading to a suppression of the field variance relative to an isolated spectator. We include a detailed treatment of the possibility that inflation may not have lasted sufficiently long for the spectator to reach its equilibrium distribution of fluctuations, finding the major constraint on such a scenario to come from isocurvature, which requires $N \gtrsim10^{10}$ e-folds. We also consider the addition of a non-minimal gravitational coupling, which extends the parameter space significantly. This model generally favors sub-GeV light dark matter ($0.1 \text{ eV} \lesssim m_S \lesssim 10 \text{ GeV}$) and in the absence of non-minimal gravity it also favors low reheating temperatures. While this mechanism can supply the entire relic abundance with sub-freeze-in portal couplings, it also naturally provides a suppressed initial condition for standard Higgs portal freeze-in.
\end{abstract}
\maketitle


\section{Introduction}
\label{Sec:Introduction}

Dark matter is often assumed to be absent at the end of inflation and 
subsequently produced through interactions with a thermal bath. This
assumption, however, is not generally justified for scalar dark matter. Any
sufficiently light scalar field $S$ develops long-wavelength quantum
fluctuations during inflation \cite{fluc}. On super-Hubble scales, these fluctuations
behave as a nearly homogeneous classical condensate and therefore provide an
irreducible primordial population before any conventional thermal production
mechanism becomes operative.

The growth of these fluctuations is stochastic in nature, and initially grows linearly in time 
during inflation,
\beq
\langle S^2 \rangle
\simeq
\left(\frac{H_I}{2\pi}\right)^2  N
=
\frac{H_I^3}{4\pi^2}(t-t_0) \, ,
\label{lingr}
\eeq
where $N$ refers to the total number of e-folds during inflation
and $H_{I}$ is the Hubble parameter during inflation.
For a quadratic spectator potential and an inflationary period of sufficiently long duration, the magnitude of the fluctuations saturates at
\beq
S_i^2 \equiv \left\langle S^2\right\rangle_{\rm end}
=
\frac{3H_{\rm end}^4}{8\pi^2m_{S,{\rm eff}}^2} \, ,
\label{Sisat}
\eeq
where $H_{\rm end}$ is the Hubble parameter at the end of inflation. The long-wavelength modes of these fluctuations satisfy the classical equation of motion for a homogeneous scalar field \cite{Linde:1983mro}.
 After
inflation, a light, weakly interacting scalar field remains overdamped until the expansion rate becomes
comparable to its mass, after which it oscillates and redshifts as
non-relativistic matter, possibly a candidate for dark matter \cite{Turner:1987vd, Peebles:1999fz, Enqvist:2014zqa, Nurmi:2015ema, Bertolami:2016ywc, Alonso-Alvarez:2018tus,ema,Markkanen:2018gcw, Tenkanen:2019aij, Cosme:2020nac,Ling:2021zlj,Basso:2022tpd,Lebedev:2022cic,Kaneta:2023kfv,Choi:2024bdn,Garcia:2025rut,Feiteira:2025rpe}. 
The late-time abundance of scalar dark matter depends on $S_i$ and therefore may retain a strong
memory of the inflationary epoch.

It is easy to see why this may be problematic for a non-interacting light scalar. If the fluctuations are saturated, that is, $S_i$ is given by Eq.~(\ref{Sisat}), then assuming that the oscillations begin in a radiation dominated universe, they begin when $H \sim m_S$ or when the temperature of the radiation is $T_{\rm osc} \sim \sqrt{m_S M_P} $, where $M_P = 2.4\times 10^{18}$~GeV is the reduced Planck mass. The energy density in $S$ today is 
$\rho_S \sim m_S^2 S_i^2 (T_0/T_{\rm osc})^3 \sim H_{\rm end}^4 T_0^3/(m_S M_P)^\frac32$, where $T_0$ is the present temperature of the cosmic microwave background (CMB). Then relative to the critical density we have
$\Omega_S h^2 \sim 4 \times 10^{27} (100~{\rm GeV}/m_S)^\frac32$, where we have taken $H_{\rm end} \sim 7 \times 10^{12}$~GeV and ignored factors of $2,~ \pi$ etc. (these will be included more rigorously in our analysis below). This is a disastrous conclusion for a stable light scalar field.

Several mechanisms can modify this conclusion. Spectator self-interactions
can affect both the stochastic distribution generated during inflation and
the subsequent evolution of the condensate \cite{Enqvist:2014zqa,Nurmi:2015ema,Markkanen:2018gcw,Cosme:2020nac,Lebedev:2022cic,Kaneta:2023kfv,Garcia:2025rut}, while direct couplings to the
inflaton can modify its effective mass and post-inflationary dynamics
\cite{Choi:2024bdn,Garcia:2025rut}. These examples suggest a more general
question: can an interaction which connects scalar dark matter to the
visible sector (in a gauge invariant manner) significantly modify this inflationary memory?

We address this question by considering the minimal Higgs portal coupling,
\begin{align}
V(S,H)
\supset &
\frac{1}{2}m_S^2S^2
+  \frac{\lambda_S}{4}S^4 +
\frac{\lambda_{HS}}{2}S^2H^\dagger H \nonumber \\
 & + m_H^2 H^\dagger H + \lambda_H (H^\dagger H)^2\, .
\label{Eq:PortalPotential}
\end{align}
Despite its simplicity, this interaction affects the spectator at different stages of its cosmological evolution \cite{Enqvist:2014zqa,Nurmi:2015ema,Lebedev:2022cic}. 

First of all, if the scalar quartic self-coupling is substantial, the effective mass squared of $S$, at large $S$, is $3 \lambda_S S^2$ and $S_i$ is significantly reduced \cite{Cosme:2020nac,Garcia:2025rut}
\beq
S_i^2 = \frac{\Gamma(\frac{3}{4})}{\Gamma(\frac{1}{4})}\sqrt{\frac{3}{2\pi^2}}\frac{\Hend^2}{\sqrt{\lambda_{S}}}\simeq 0.132 \frac{H_{\rm end}^2}{\sqrt{\lambda_S}},
\label{Eq:Sendquartic}
\eeq
The final relic density in this case is roughly $\Omega_S h^2/0.12 \sim 16 \lambda_S^{-\frac58} (m_S/100~{\rm GeV})$ \cite{Garcia:2025rut}. Clearly the effective mass of $S$ plays a crucial role in the determination of the final relic density.  We note that even if $\lambda_S$ is absent at tree level or tuned to a very small value, Higgs loops will always radiatively generate a
quartic self-interaction 
\beq
\lambda_S^{\rm rad}
\simeq
\kappa\frac{\lambda_{HS}^2}{16\pi^2}\, ,
\label{Eq:RadiativeQuarticIntro}
\eeq
where $\kappa$ is a coefficient of order unity. Because the presence of the portal coupling will always induce the quartic self-coupling, we have included it in the potential even though we will almost always take the spectator's bare self-interaction term $\lambda_{S,0}=0$ throughout this work, such that $\lambda_S=\lambda_{S}^{\rm rad}$. 

In addition, 
once a Standard Model thermal bath is established, the portal
generates a thermal contribution to the spectator mass \cite{Enqvist:2014zqa},
\beq
m_{S,\rm th}^2(T)
=
c_S T^2\, ,
\label{Eq:ThermalMassIntro}
\eeq
with\footnote{In principle, $c_S$ could be written as $\frac{1}{12} \left( 2 \lambda_{HS} + 3 \lambda_S \right)$, however, in this work we focus on the sub-feeble coupling regime such that there is no significant population of spectators in the thermal bath to provide this loop correction.}
\beq
c_S = \frac{1}{6}  \lambda_{HS} \, .
\label{Eq:cs}
\eeq
This expression corresponds to the leading high-temperature result in the
electroweak-symmetric phase and should not be extrapolated below the
electroweak scale. Below the electroweak scale, the Higgs vacuum expectation value (VEV) will generate a contribution,
\beq
m_{S,v}^2 = \frac12 \lambda_{HS} v^2 \, ,
\label{vcontr}
\eeq
where we take $v=246$~GeV. Note that unlike the self-coupling contribution to the effective mass, the thermal corrections do not affect the initial fluctuation, $S_i$, coming out of inflation. However, 
the post-inflationary cosmological evolution of the primordial condensate is affected by the total effective mass \footnote{Note that the 2nd and 4th terms in this expression do not typically contribute to $m_{S,\rm eff}^2$ at the same time. At high temperatures, $c_S T^2$ contributes, and while $\langle H^\dagger H \rangle \ne 0$ initially, the Higgs condensate decays away very rapidly since it interacts with the thermal bath with a relatively strong coupling. However the 4th term does play a role below the electroweak transition when $\langle H^\dagger H \rangle = v^2/2$ but at that point, the Higgs drops out of the thermal bath and we lose the contribution from the 2nd term.}
\beq
m_{S,\rm eff}^2(S,T, H)
=
m_S^2
+
c_S T^2 +
3\lambda_S S^2 + \lambda_{HS} \langle H^\dagger H \rangle \, .
\label{Eq:EffectiveMassIntro}
\eeq
As mentioned above, $\lambda_S$ includes the one-loop contributions from $\lambda^{\rm rad}_S$, and we will typically take $\lambda_S = \lambda_{S}^{\rm rad}$ throughout this work since it is the minimal unavoidable contribution to the spectator's self-interaction in this model.
Equation~(\ref{Eq:EffectiveMassIntro}) encapsulates the central physical
picture of this work. The bare, radiative, and thermal masses have different
cosmological dependences and can successively dominate the spectator
dynamics. The radiative mass can control the stochastic fluctuations
generated during inflation and/or the initial onset of oscillations, whereas the thermal mass can dominate during reheating or radiation domination and aid in diluting the primordial condensate. At sufficiently late times, the bare mass and the portal mass eventually take over and the surviving condensate
behaves as cold dark matter. Note that all terms in Eq.~(\ref{Eq:EffectiveMassIntro}) except the bare mass will fundamentally depend on $\lambda_{HS}$. Of course, dominance of a given contribution
to the effective mass does not necessarily imply that the spectator is
already oscillating. Rather, the onset of oscillations must independently satisfy
\beq
m_{S,\rm eff}(T_{\rm osc})\simeq \frac32 (1+w) H(T_{\rm osc}), 
\label{Eq:QuadraticOscillation}
\eeq  where $w = 0$ (1/3) is the equation of state parameter for matter (radiation) domination.

We will see that the two additional contributions to the effective spectator mass in Eq.~(\ref{Eq:EffectiveMassIntro}) induced by the Higgs portal which will be present at high temperatures (namely, the thermal mass term and the radiative quartic term) can fundamentally alter the evolution of the spectator scalar condensate in a way that is favorable for dark matter production, even for extremely feeble portal couplings ($10^{-28} \lesssim \lambda_{HS}\lesssim 10^{-9}$). Specifically, when oscillations occur in either the radiative quartic or thermal mass regime, the condensate's energy density will efficiently redshift as $\rho_S\propto a^{-k}$ with $k>3$, rather than the milder matter-like scaling $\rho_S\propto a^{-3}$. This favors efficient and early suppression of scalar dark matter abundance compared with isolated spectators. The underlying mechanisms are different for the two terms, despite each originating from the same term in the Lagrangian. In the quartic regime the altered evolution follows from the equation of state of the oscillating
field, whereas in the thermal regime it results from the time-dependent (decreasing) oscillation frequency $m_{S,\rm eff}\propto T\propto a^{-1}$. \textit{The Higgs
portal can therefore suppress a large fraction of the initial inflationary abundance simply due to early quartic- or thermal mass-induced oscillations,
without thermalizing or destroying the spectator condensate}. This is the central mechanism we identify and explore in this work.

Higgs portal interactions with light spectators have previously 
been studied in various contexts. For instance, the same $\mathbb{Z}_2$ symmetric Higgs portal model we study here was previously studied in Refs. \cite{Enqvist:2014zqa, Nurmi:2015ema}. There are several 
important differences between our work and these earlier works. First, the authors of  Refs. \cite{Enqvist:2014zqa,Nurmi:2015ema} comment that the spectator condensate will not ``feel the thermal bath" for $\lambda_{HS}\lesssim 10^{-7}$. As a consequence they do not include a thermal mass for the spectator
for smaller Higgs portal interactions. While it is true that the 
condensate may be preserved for $\lambda_{HS} \lesssim 10^{-7}$, and need not 
evaporate to physical quanta (see further details in Appendix \ref{app:evap}.), the thermal mass provided by Eq.~(\ref{Eq:ThermalMassIntro}) may be non-negligible and alter the final spectator abundance. Indeed,  we will see throughout this work that the thermal mass induced by the coupling to the Higgs boson is often extremely important for determining the cosmological evolution of the spectator condensate even for $\lambda_{HS}$ much smaller than $10^{-7}$. 

Second, the authors in Refs. \cite{Enqvist:2014zqa,Nurmi:2015ema} consider a prolonged 
inflationary epoch in which the light spectator scalar reaches its 
equilibrium distribution. While we also consider this as a sub-case in this work, as we will review in Section \ref{Sec:InflationaryMemory}, it would often require more than $10^{20}$ e-folds of inflation to reach this equilibrium 
distribution depending on the spectator mass, self-coupling, and 
inflationary scale. This motivated us to additionally consider scenarios where inflation is shorter than this, such that the light 
scalar fluctuations do not reach their equilibrium distribution. We note however, that the duration cannot be arbitrarily small if light stable spectators are present due to constraints from isocurvature fluctuations \cite{Kainulainen:2016vzv,Heikinheimo:2016yds,Garcia:2022vwm,Garcia:2023awt,Garcia:2025wmu,Garcia:2025rut}. Lastly, these prior works did not study regimes where the stochastic equilibrium distribution of the spectator fluctuations was dictated by the Higgs portal interaction. In particular, in Refs. \cite{Enqvist:2014zqa,Nurmi:2015ema}, the authors assume $\lambda_{HS}$ is small and expand perturbatively, which amounts to taking $\lambda_{HS}\ll \sqrt{\lambda_S \lambda_H}$. Upon doing so, they find a stochastic equilibrium distribution for the spectator which is set by $\lambda_S$. In our work, we will consider the simple case where the bare self-coupling is absent ($\lambda_{S,0}=0$), such that the self-interaction is the minimal loop-induced term. In this case, we will see that $\lambda_{HS}$ will instead dictate the stochastic equilibrium distribution of the spectator.

A related mechanism to the one 
we describe here was studied in Refs. \cite{Batell:2022qvr, Batell:2026sml}. In these works, the authors consider a linear 
Higgs portal interaction of the form $\mathcal{L}\supset A S H^\dagger H$, rather than 
the quadratic interaction we use above. In their work, they found 
that this linear interaction induces a thermal misalignment of the 
scalar, which leads to a suppression of isocurvature fluctuations. 
In our work, we have a $\mathbb{Z}_2$ symmetry which remains 
present after the appearance of the Standard Model bath, such that 
there is no misalignment and the minimum of the spectator's 
potential remains at $S=0$ during inflation, reheating, and radiation domination. 

As we have noted already, 
the dramatic overproduction of 
a weak scale scalar without self interactions is largely due to the initial amplitude of the scalar condensate given by Eq.~(\ref{Sisat}). This conclusion also assumes instantaneous reheating.  That is, the reheating temperature is maximal, with $\Trh \simeq (H M_P)^{1/2}$. In this case the oscillations of the condensate always begin after reheating. If this assumption is dropped, oscillations instead may begin during reheating and the correct relic density may be achieved for relatively low reheating temperatures (100 GeV) and high scalar masses ($10^{12}$~GeV) \cite{Cosme:2020nac,Garcia:2025rut}.  

The overproduction of spectators also
assumes that the duration of inflation is sufficiently long.   Indeed a ``short" duration of inflation was used to resolve the overproduction of the baryon asymmetry \cite{Ellis:1987rw} in Affleck-Dine baryogenesis \cite{Affleck:1984fy}. However a short duration of inflation, even if justified, may only be compatible with a sub-eV spectator mass \cite{Feiteira:2025rpe}. In addition, as we will see, there is a lower bound on the duration of inflation coming from isocurvature fluctuations. 

As noted above, a resolution of the scalar overproduction is possible via a 
self coupling (quartic term) of order 1.  For instantaneous reheating, this allows spectator dark matter masses of order 6 GeV.  If we further drop the assumption of instantaneous reheating, then a large self-coupling leads to the correct relic density for a wide range of scalar masses and reheating temperatures \cite{Cosme:2020nac,Garcia:2025rut}.
There are also solutions to the overproduction problem for very small quartic couplings which satisfy the isocurvature constraint \cite{Garcia:2025rut}.

Here, we show that a Higgs portal coupling to the spectator can significantly affect the resulting relic abundance of spectator dark matter. For a large portal coupling, the condensate is destroyed, and if it is sufficiently large, $S$ is no longer a spectator, but rather a WIMP.  For a small but still sufficiently large value of the portal coupling ($10^{-11} \lesssim \lambda_{HS}\lesssim 10^{-7}$, depending on $m_S$ and $T_{\rm RH}$),
the relic abundance will often be controlled by freeze-in \cite{fimp,Bernal:2017kxu}. 
However, as we show, even a very small portal coupling (even more feeble than those associated with freeze-in, namely $\lambda_{HS}\lesssim 10^{-10}$)
plays a qualitatively different role from conventional freeze-out
or freeze-in: rather than simply producing dark matter, it controls how much
of a population generated during inflation survives until today. The relic
abundance consequently establishes a connection between the microscopic
Higgs portal interaction and the energy scale at the end of inflation.

In what follows, we will conduct a systematic investigation of spectator dark matter coupled to the visible sector through a feeble (or sub-feeble) Higgs portal interaction. To that end, we consider the following set of parameters for the model
\beq
m_S, \, \lambda_S, \,  \lambda_{HS}, \, \Trh, \, N \, .
\eeq
The first three of these appear directly in the scalar potential for $S$. Although we quote some results for non-vanishing $\lambda_S$, we will for the most part assume that the tree-level value is small so that $\lambda_S \propto \lambda_{HS}^2$ as in Eq.~(\ref{Eq:RadiativeQuarticIntro}).
The reheating temperature if it is not maximal can affect the evolution of the condensate and hence the final relic density. 
The parameter $N$ refers to the total number of inflationary e-folds
as in Eq.~(\ref{lingr}).  As we discuss in the next section, for a model of inflation such as the Starobinsky model \cite{Staro}, $N$ can be virtually infinite as the plateau in the inflationary potential admits eternal inflation \cite{Vilenkin:1983xq,Linde:1986fd}. However, even very small deformations of this potential \cite{Antoniadis:2025pfa} lead to a finite value of $N$ essentially independent of initial conditions. 

In the next section, we begin by discussing the inflationary sector that determines $\Trh$ and $N$ as well as the constraints from isocurvature fluctuations. In Section \ref{Sec:InflationaryMemory} we review the known results for the spectator governed by both quadratic and quartic terms when oscillations occur before and after reheating. There we also introduce results for a duration $N < N_{\rm eq}$ less than the equilibrium value. In Section \ref{Sec:EffectiveMassHierarchy}, we outline four spectator oscillation scenarios which are distinguished by the hierarchy of terms in the effective mass, and we briefly describe the joint stochastic evolution of the $S$ and $h$ during inflation. Our main results are given in Section \ref{Sec:ThermalReshaping}, where we concentrate on the effect of the quartic and thermal mass contributions, and we describe the striking effects of adding a non-minimal gravitational coupling. Our conclusions are summarized in Section \ref{Sec:Conclusion}.

\section{Inflation and Reheating}
\label{Sec:InfReh}

In this section, we consider the origin of the scalar fluctuations leading to spectator dark matter. As a working example, we consider the Starobinsky potential \cite{Staro}
\begin{equation}
\label{treestaro}
V \; = \; \frac34 M^2 M_P^2\left[ 1 - e^{-\sqrt{\frac23} \frac{\phi}{M_P} }\right]^2 \, .
\end{equation}
 In its original formulation, this potential is derived from a theory of gravity with an $R^2$ correction to the action. Specifically, $R - \frac{1}{6M^2} R^2$, accounting for the mass scale, $M$, in (\ref{treestaro}) and equal to the inflaton mass. CMB observables can be derived from the slow-roll parameters
\begin{equation}
\epsilon \; = \; \frac{M_P^2}{2} \left( \frac{V^\prime}{V} \right)^2 \; , \; \eta = M_P^2 \frac{V^{\prime \prime}}{V} \, ,
\end{equation}
where prime denotes a derivative with respect to $\phi$. For example, the amplitude of the scalar anisotropy spectrum  $A^*_s = 2.1 \times 10^{-9}$ \cite{Planck,Plancki} can be used to fix the inflationary mass scale, $M$, using 
\beq
A^*_s   =   \frac{V(\phi_*)}{24\pi^2 \epsilon_* M_P^4} =  \frac{3 M^2}{8\pi^2 M_P^2} \sinh^4 \left(\frac{\phi_*}{\sqrt{6}M_P} \right)\, . \label{As2}
\eeq
The subscript $*$ denotes field values chosen to correspond to a pivot scale, often taken to be  $k_* = 0.05$~Mpc$^{-1}$.
This is fixed by the number of $e$-folds between
the horizon exit at the pivot scale and the end of inflation: 
\begin{equation}
N_*  \; \simeq \;  - \int^{\phi_{\rm{end}}}_{\phi_*} \frac{1}{\sqrt{2 \epsilon}} \frac{d \phi}{M_P} \, .
\label{e-folds}
\end{equation}
The end of inflation is defined when $\ddot{a}=0$, where $a$ is the cosmological scale factor. This occurs when the inflaton field value is $\phi = \phi_{\rm end}$.
The field value at the pivot scale is $\phi = \phi_*$. For the Starobinsky potential $\phi_{\rm end} \simeq 0.63 M_P$ and $\phi_* = 5.35 M_P$ for $N_* = 55$ (see e.g., \cite{EGOV}). 

For this value of $\phi_*$, from (\ref{As2}), the inflaton mass is $M=3 \times 10^{13}$~GeV. \footnote{Note that there is a small (log) dependence in $N_*$ on the reheating temperature which affects $\phi_*$, the normalization of the inflaton mass as well as $\rhoend$ and $\Hend$. } As a consequence, at the value $\phi = \phi_{\rm end}$, the energy density of the inflaton condensate is $\rho_{\rm end}^\frac14 = 5.6 \times 10^{15}$~GeV and the value of the Hubble parameter is $H_{\rm end} = 7.4 \times 10^{12}$~GeV.

The total number of inflationary e-foldings is dependent on the initial field value, $\phi_0$
\begin{equation}
N  \; \simeq \;  - \int^{\phi_{\rm{end}}}_{\phi_0} \frac{1}{\sqrt{2 \epsilon}} \frac{d \phi}{M_P} \, .
\label{e-foldst}
\end{equation}
For the Starobinsky potential, the total number of e-folds
increases exponentially with increasing $\phi_0$, $N \simeq \frac34 e^{\sqrt{2/3} \frac{\phi_0}{M_P}}$ for large $\phi_0$.
Indeed for large $\phi_0$, the potential is very flat and quantum fluctuations may push the inflaton to smaller or larger field values, leading to eternal inflation \cite{Vilenkin:1983xq,Linde:1986fd}. 
Our patch of the Universe derives from a region where the inflaton value was sufficiently small that the equation of motion wins over quantum fluctuations, allowing the inflaton to evolve towards the minimum of the potential. 
During inflation, quantum fluctuations of all light scalar fields (with approximately constant effective mass) grow as 
\beq
\langle S^2 \rangle =\frac{3H_{\rm I}^4}{8\pi^2m_{S,{\rm eff}}^2} \left(1 - e^{-\frac{2 m_{S,\rm{eff}}^2}{3 H_{\rm I}}(t-t_0)} \right) \, ,
\eeq
where $H_{\rm I}^2 \simeq V(\phi_0)/3 M_P^2$ is the value of the Hubble parameter during inflation.
Initially (for small $H_{\rm I} (t-t_0)$), these fluctuations grow linearly in time, as given by Eq.~(\ref{lingr}). If the duration of inflation is sufficiently long, 
the fluctuations saturate the value given by Eq.~(\ref{Sisat}). Note that we take the initial value of $S_i$ to be determined by $H_{\rm end}$ rather than $H_{\rm I}$ \cite{Garcia:2025rut}. 

As we will see below, the asymptotic value for $S_i^2$ requires an extremely large value of $N$. 
In addition, a stringent lower bound on the value of $N$ can be derived from isocurvature fluctuations when a stable light spectator field is assumed to make up all of the dark matter. The isocurvature power spectrum 
is given by \footnote{We thank Marcos Garcia for providing an updated expression for the power spectrum found in  \cite{Garcia:2025wmu}. }
\beq
{\cal P}_S \propto \Lambda (1-e^{-\Lambda(N-N_*)}) e^{-2 \Lambda N_*}(1 - e^{-\Lambda N})^{-2} \sim \frac{N-N_*}{N^2} \, ,
\eeq
 where $\Lambda = 2 m_{S,{\rm eff}}^2/3 \Hend^2$ and the last term shows 
the leading term in the expansion for small $\Lambda$. Thus for $\Lambda N \ll 1$ and $N \gg N_*$, ${\cal P}_S \sim 1/N$ and the limit from Planck \cite{Plancki} gives ${\cal P}_S \lesssim 8.3 \times 10^{-11}$ or 
\beq
N \gtrsim N_{\rm iso} \simeq  10^{10}\,. 
\label{Eq:IsocurvatureN}
\eeq

After the period of exponential expansion, at $\phi < \phi_{\rm end}$ or when $a > a_{\rm end}$ the inflaton begins a series of oscillations about the minimum of the potential. Reheating occurs when inflaton decays become rapid compared to the expansion rate which is governed by the inflaton condensate. At the onset of inflaton oscillations, the first few decays lead to a radiation bath (here assumed to be thermal) which quickly reaches a maximum temperature \cite{Giudice:2000ex, Ellis:2015jpg} at $a = a_{\rm max} = (8/3)^{2/5}a_{\rm end}$, such that
\beq
\alpha_{\rm max} \Tmax^4 = \frac{\sqrt{3}}{4}\left(\frac38\right)^\frac35 \rhoend^\frac12 \Gamma_\phi M_P \, ,
\eeq
 where $\alpha_{\rm max} = g_*(a_{\rm max}) \pi^2/30$ and $\Gamma_{\phi}$ is the inflaton decay rate. The inflaton energy density at the end of inflation is $\rho_{\rm end} = \frac32 V(\phi_{\rm end})$ and $g_*(a)$ is the number of relativistic degrees of freedom at $a$. Subsequently, for $a > a_{\rm max}$, the temperature drops as
$T \propto a^{-3/8}$ \cite{Scherrer:1984fd,Giudice:2000ex,GKMO1,GKMO2} until reheating is achieved (defined by $\rho_\phi (\arh) = \rho_{\rm R}(\arh)$). The reheating temperature is also determined by $\Gamma_\phi$, namely
\beq
\alpha_{\rm RH} \Trh^4 = \frac{12}{25} \Gamma_{\phi}^2 M_P^2 \, ,
\eeq
where now $\alpha_{\rm RH} = g_*(\arh) \pi^2/30$.

\section{The inflationary memory of spectator dark matter}
\label{Sec:InflationaryMemory}

We begin by distinguishing two distinct cases for the duration of inflation. First, in Sec. \ref{Sec:Equilibrium} we consider an inflationary epoch which is sufficiently long such that a light spectator will reach its equilibrium distribution of fluctuations. This is commonly used to obtain initial conditions for spectator scalars at the end of inflation. 
 In Sec. \ref{Sec:ShortInflation} we consider the alternative possibility that inflation was not sufficiently long for the condensate to reach its equilibrium distribution. In this case, the spectator fluctuations behave essentially like a random walk if $N \ll H^2/m_{S,\rm eff}^2$, as discussed above. Throughout this section, we consider isolated spectators in the absence of a Higgs portal interaction, which will provide us with useful stepping stones when we add the Higgs portal in Sections \ref{Sec:EffectiveMassHierarchy} and \ref{Sec:ThermalReshaping}.

\subsection{Stochastic equilibrium: long inflation}\label{Sec:Equilibrium}

\subsubsection{Quadratic domination}
\label{Sec:QuadEq}

A light spectator scalar field inevitably acquires long-wavelength
fluctuations during inflation, which behave on super-Hubble scales as a
nearly homogeneous classical condensate
\cite{Enqvist:2014zqa,Nurmi:2015ema,Markkanen:2018gcw,
Tenkanen:2019aij,Choi:2024bdn,Garcia:2025rut}.
For an isolated light spectator with a quadratic potential and a sufficiently long period of inflation, stochastic growth of fluctuations continues until $\langle S^2 \rangle$ reaches its equilibrium value given in Eq.~(\ref{Sisat}) with a corresponding energy density given by
\beq
\rho_S^{\rm end}
=\frac12 m_S^2
\left\langle S^2\right\rangle_{\rm end}=
\frac{3H_{\rm end}^4}{16\pi^2}\, .
\label{Eq:QuadraticInitialConditions}
\eeq
The primordial energy density is therefore independent of $m_S$ and
constitutes an irreducible inflationary abundance.  The stochastic equilibrium expressions above, however, often require extremely long inflationary periods to realize. In particular, for the standard quadratic spectator, the number of e-folds $N$ required for equilibration is
\beq
N \simeq \mathcal{O}\left(\frac{H^2}{m_S^2}\right) \, .
\label{bigN}
\eeq 
For example, for $H_{\rm inf}\sim 10^{13}$~GeV and $m_S=100$~GeV, we have $N\sim10^{22}$.  One might ask whether such an extremely long period of inflation is natural, a question we set aside for the remainder of this section. We return to this issue in the following section, where we examine in detail the alternative scenario of a shorter inflationary epoch.

To illustrate the resulting overproduction problem from stochastic equilibrium, consider first the
purely quadratic evolution. The spectator
remains overdamped until the bare mass exceeds the Hubble rate (specifically, such that Eq.~(\ref{Eq:QuadraticOscillation}) is satisfied). At the onset of oscillations, the energy density in $S$ is given by Eq.~(\ref{Eq:QuadraticInitialConditions}).
Afterwards the energy density redshifts as $\rho_S\propto a^{-3}$.
If $\Tosc > \Trh$, the Hubble parameter is dominated by inflaton oscillations,
\beq
H = \frac{\rho_\phi^\frac12}{\sqrt{3}M_P} = \sqrt{\frac{\alpha}{3}} \frac{T^4}{\Trh^2 M_P} \, ,
\label{htinfl}
\eeq
where we used $\rho_{\rm RH} = \alpha_{\rm RH} \Trh^4 = \rhoend (\aend/\arh)^3$. 
In this case, 
\beq
\alpha \Tosc^4 = 2 \sqrt{\frac{\alpha}{3}} m_S \Trh^2 M_P
\eeq
and
the condition $\Tosc > \Trh$ corresponds to \cite{Garcia:2025rut}
\beq
m_S > \frac{\sqrt{3 \alpha}}{2} \frac{\Trh^2}{M_P} \simeq 210 \left( \frac{\Trh}{10^{10}~{\rm GeV}} \right)^2~\rm GeV\,,
\label{msbeforereh}
\eeq
assuming $g_* = 427/4$.
The relic spectator density is then \cite{ema2, Cosme:2020nac, Choi:2024bdn,Garcia:2025rut}
\beq
\frac{\Omega_S h^2}{0.12} \; \simeq \; 1.2 \frac{\Trh}{100~{\rm GeV}}\left(\frac{10^{12}~\rm GeV}{m_S}\right)^2
\left(\frac{\Hend}{7.4\times 10^{12}~{\rm GeV}} \right)^4  \,.
\label{Eq:casek2}
\eeq
Thus if oscillations take place during reheating, the relic density is acceptable for low reheating temperature and high spectator masses. 

If Eq.~(\ref{msbeforereh}) is not satisfied, then oscillations begin after reheating. The Hubble rate is now dominated by the radiation density and
\beq
H(T)=
\sqrt{\frac{\pi^2g_*(T)}{90}}\,
\frac{T^2}{M_P}\, .
\eeq
Redshifting Eq.~(\ref{Eq:QuadraticInitialConditions}) to the present gives
\begin{align}
\frac{\Omega_Sh^2}{0.12}
& \simeq
1.3 \times 10^{28}
\left(\frac{100~\rm GeV}{m_S}\right)^\frac32
\left( \frac{\Hend}{7.4\times 10^{12}{\rm GeV}} \right)^4 \nonumber \\  &\times\left(\frac{106.75}{g_* (T_{\rm osc})}\right)^{1/4}.
\label{Eq:OmegaQuadraticb}
\end{align}
Thus, for a weak-scale spectator, with $\Tosc < \Trh$, the inflationary condensate
overcloses the Universe by nearly thirty orders of magnitude. Moreover,
$\Omega_Sh^2\propto m_S^{-3/2}$, so the problem becomes increasingly severe
at smaller masses. Even at $m_S = 10^{12}$~GeV, $\Omega_S h^2$ is 13 orders of magnitude too large. Note that this result also corresponds to the case of instantaneous reheating, when $\Trh^2 \simeq \Hend M_P$.

\subsubsection{Quartic domination}

It is well known that the above result for an isolated quadratic spectator changes rather dramatically when a quartic self-interaction term is added. Both the equilibrium variance and the final relic abundance are different from the quadratic case. In particular, the quartic term will dominate the equilibrium variance when $m_S^2 \ll 3\lambda_S \langle S^2\rangle_{\rm end}$, or equivalently
\beq
m_S^2 < 3 (.132) \sqrt{\lambda_S} \Hend^2 \, .
\eeq
 In this case, the equilibrium variance is given by Eq.~(\ref{Eq:Sendquartic}). 

The final relic abundance when both quadratic and quartic terms are present depends on the relative contributions of these terms to the energy density and effective mass, which will determine the onset of oscillations and the subsequent redshifting of the energy density. If the self-interaction induces the onset of oscillations, we must also then determine the point at which the bare mass term overtakes the self-interaction (at $a = a_{\rm m}$).
In other words, we must determine the relative size of the scale factors $a_m$, $a_{\rm osc}$, and $\arh$. We assume that all of these are greater than $\aend$. 
For example, when the bare mass term dominates after oscillations begin, ($a_{\rm osc} < a_{\rm m} < \arh$), we have \cite{Garcia:2025rut}
\beq
\frac{\Omega_S h^2}{0.12}\simeq \lambda_S^{-\frac{3}{4}}
 \left(\frac{m_S}{10^{12}~\rm GeV} \right)
 \left(\frac{\Trh}{14~\rm TeV}\right) \left(\frac{\Hend}{7.4\times 10^{12}~{\rm GeV}} \right) \, .
\label{Eq:k2p4}
\eeq
Independently, the requirement that the quartic-to-quadratic transition occur before the end of reheating, $a_m<\arh$, imposes
\beq
m_S
\gtrsim
0.02~{\rm GeV}\,
\lambda_S^{\frac{1}{12}}
\left(
\frac{H_{\rm end}}{7.4\times10^{12}~{\rm GeV}}
\right)^{\frac13}
\left(
\frac{T_{\rm RH}}{40~{\rm TeV}}
\right)^{\frac43}
\, .
\label{Eq:ms1}
\eeq
\noindent

While the correct relic density is found for similar $m_S$ and $\Trh$ for $\lambda_S \sim 1$ (compared to Eq.~(\ref{Eq:casek2})), the dependence on the mass is very different. With the
quartic coupling, increasing the mass shortens the period between $a_{\rm osc}$ and $a_{\rm m}$ allowing less time for dilution and increasing the
relic density. Combining condition (\ref{Eq:ms1}) with the requirement $\Omega_S h^2=0.12$ from Eq.~(\ref{Eq:k2p4}) yields an upper bound on the reheating 
temperature,
\beq
T_{\rm RH}
\lesssim
1.9\times10^{10}~{\rm GeV}\,
\lambda_S^{\frac27}
\left(
\frac{7.4\times10^{12}~{\rm GeV}}{H_{\rm end}}
\right)^{\frac47}
\, .
\eeq
For $a_{\rm osc} < \arh <  a_{\rm m}$, we require
 $m_S \lesssim 0.02 M_P \lambda_S^{\frac{1}{12}} (\Trh/M_P)^\frac43$, and the expression for the relic density in (\ref{Eq:k2p4}) still holds.

For sufficiently large reheating temperatures such that
$a_{\rm RH}<a_{\rm osc}<a_{\rm m}$, the final relic abundance is
\cite{Garcia:2025rut}\footnote{Throughout this analysis, we restrict
ourselves to spectator masses $m_S<H_{\rm end}$, for which inflationary
fluctuations of $S$ are not exponentially suppressed.}
\beq
\frac{\Omega_S h^2}{0.12}
\simeq
\lambda_S^{-\frac{5}{8}}
\left(\frac{m_S}{6.4~{\rm GeV}}\right)
\left(\frac{\Hend}{7.4\times10^{12}~{\rm GeV}}\right)^{\frac32}
\, .
\label{k4p4}
\eeq

We note that the benchmark $\lambda_S={\cal O}(1)$ lies in the
large self-coupling branch compatible with the isocurvature constraint
\cite{Garcia:2025rut},
whereas a second allowed branch exists for extremely small
quartic couplings,
\beq
\lambda_S
\lesssim
\lambda_{S,\rm iso}
\simeq 2\times 10^{-21}
\, .
\label{Eq:IsocurvatureQuartic}
\eeq
We will return to these
isocurvature constraints later in the discussion.

The expression (\ref{k4p4}) is valid provided reheating is completed before the onset of quartic oscillations,
\bea
\Trh>\Tosc
&&
\simeq
0.3\,\lambda_S^{\frac18}
\left(\Hend M_P\right)^{\frac12}
\\
&&
\simeq
1.3\times10^{15}~{\rm GeV}\,
\lambda_S^{\frac18}
\left(\frac{\Hend}{7.4\times10^{12}~{\rm GeV}}\right)^{\frac12}
\, .
\nonumber
\eea
In this regime, the observed relic abundance can therefore be obtained for spectator masses around the weak scale for $\lambda_S={\cal O}(1)$, though it would require a rather large reheating temperature.

These examples illustrate that self-interactions are one way to overcome the catastrophic overproduction problem associated with the isolated quadratic spectator.  Note that this solution has additional constraints from the production of isocurvature fluctuations \cite{Garcia:2022vwm,Garcia:2023awt,Garcia:2025wmu,Garcia:2025rut} and fragmentation \cite{Garcia:2025rut}. These require quartic couplings either of order 1, or $\lambda_S \lesssim 2 \times 10^{-21}$. Furthermore, avoiding a second phase of inflation requires $\lambda_S \gtrsim 10^{-28}$ \cite{Garcia:2025rut}.
The spectator density is also under control if $S$ couples to the inflaton \cite{Choi:2024bdn,Garcia:2025rut}, though we do not consider that possibility here. In this work, we ask whether a minimal and very feeble Higgs portal interaction might be an independent, alternative mechanism to overcome the overabundance problem.



\subsection{Finite $N$: shorter inflation}
\label{Sec:ShortInflation}

\subsubsection{Quadratic domination}

The above results demonstrating a catastrophic overabundance for an isolated quadratic spectator used an initial condition for $\langle S^2 \rangle_{\rm end}$ which relied on its equilibrium distribution. However, reaching this equilibrium distribution often would require an extremely large number of inflationary e-folds, given by Eq.~(\ref{bigN}). 
As argued in the previous section, even slight deformations of the Starobinsky potential can yield a value of $N$ far smaller than that in Eq.~(\ref{bigN}).
If the duration of inflation is such that the fluctuations do not reach their equilibrium value, 
they instead retain their random-walk behavior  and 
\beq
\langle S^2 \rangle_{\rm end} \simeq \frac{N H_{\rm end}^2}{4 \pi^2} \, ,
\label{Eq:S2endFiniteN}
\eeq 
for $N \ll H^2/m_{S,\rm eff}^2$ \footnote{Note that Eq.~(\ref{Eq:S2endFiniteN}) assumes that the pre-inflationary initial condition for the fluctuations is negligible compared to the random walk contribution. We make this assumption throughout this work for simplicity. For a thorough treatment and discussion of the pre-inflationary initial condition, see Ref. \cite{Feiteira:2025rpe}.}. As a consequence, the energy density at the onset of oscillations is 
\beq
\rho_S^{\rm end}
\simeq
N \frac{H_{\rm end}^2 m_S^2}{8\pi^2}\, ,
\label{Eq:QuadraticInitialConditionsN}
\eeq
and the resulting final relic density can be obtained from the preceding results in Section \ref{Sec:QuadEq} by rescaling the stochastic equilibrium value of $\rho_S^{\rm end}$ in Eq.~(\ref{Eq:casek2}) by a factor $2N m_S^2/(3\Hend^2)$.

We recall from Eq.~(\ref{Eq:IsocurvatureN}) and the surrounding discussion that in the finite $N$ (shorter inflation) scenario, there is a major constraint on the number of inflationary e-folds imposed by isocurvature fluctuations. Therefore, in order for the spectator to both remain in the finite-$N$, non-equilibrium regime and satisfy the isocurvature constraint, we require
\beq
N_{\rm iso}
\lesssim N
\ll N_{\rm eq}
\sim \frac{3H_{\rm end}^2}{2m_S^2}
\, .
\label{Nisoquad}
\eeq
The applicable mass range therefore for $N < N_{\rm eq}$ becomes
\beq
m_S
\lesssim
\sqrt{\frac32}
\frac{H_{\rm end}}{\sqrt{N_{\rm iso}}}
\simeq
9\times10^7~{\rm GeV}
\left(\frac{H_{\rm end}}{7.4\times10^{12}~{\rm GeV}}\right)
\, .
\label{Eq:miso}
\eeq
\noindent
For larger masses, the isocurvature constraint forces us to values of $N$ equal to the equilibrium value. 
We stress, however, that Eq.~(\ref{Eq:miso}) applies only when the
inflationary dynamics are dominated by the quadratic term. 

While a shorter duration of inflation can lead to a much smaller amplitude for the stochastic spectator fluctuations at the end of inflation (as evident from Eq.~(\ref{Eq:S2endFiniteN})), it is not by itself a cure-all for the overproduction problem especially since the isocurvature constraint is so severe. 
For example, in the case where oscillations begin during non-instantaneous reheating (namely $T_{\rm osc}>T_{\rm RH}$, corresponding to the constraint in Eq.~(\ref{msbeforereh})), we find the following result for the relic abundance given the finite $N$ initial conditions for the spectator fluctuations in Eq.~(\ref{Eq:S2endFiniteN})
\begin{align}
\frac{\Omega_S h^2}{0.12}
\simeq
6\times10^{3}
\left(\frac{N}{10^{10}}\right)
\left(\frac{H_{\rm end}}{7.4\times10^{12}\,{\rm GeV}}\right)^2
\left(\frac{T_{\rm RH}}{4 \,{\rm MeV}}\right)
\, .
\label{Eq:FiniteNreh}
\end{align}
Interestingly, in contrast to both the quadratic and quartic stochastic equilibrium cases in Eqs.~(\ref{Eq:casek2}) and (\ref{Eq:k2p4}), respectively, the relic abundance in Eq.~(\ref{Eq:FiniteNreh}) is \emph{independent of the spectator mass}. This cancellation can be 
understood from the finite-$N$ initial condition: since $\langle S^2\rangle_{\rm end}\propto N H_{\rm end}^2$ is independent of $m_S$, the 
initial spectator energy density scales as $\rho_S^{\rm end}\propto m_S^2$, 
which precisely compensates the additional dilution associated with the earlier onset of oscillations for larger $m_S$.

However from the above expression, we see that finite $N$ inflation in combination with non-instantaneous reheating is still insufficient for obtaining the correct relic density, as we have already normalized the reheating temperature to the BBN limit $\Trh > 4$~MeV \cite{tr4} and the isocurvature bound on $N\sim 10^{10}$.  
Thus, even under the most favorable assumptions, the isolated quadratic
spectator with shorter inflation is still overproduced by about four orders of magnitude.
Therefore, there is no overlap between the
relic density and isocurvature constraints for an isolated quadratic
spectator whose oscillations begin during reheating. This provides an
additional motivation for the Higgs portal-induced mechanism
considered below.

Note that some features of this result, such as its independence on the 
spectator mass, were observed in Ref.~\cite{Feiteira:2025rpe}. In this work, the authors find a required size of the spectator fluctuations compatible with the correct abundance (for a fixed choice of $T_{\rm RH}$) which can be converted into a required value of $N$ (assuming a negligible pre-inflationary initial condition). However, as we have described in detail above, the required value of $N$ needed to be consistent with the correct abundance is in severe conflict with the isocurvature constraints.

For larger reheating temperatures such that $T_{\rm osc} < T_{\rm RH}$, the parameter space is greatly restricted.
Replacing the stochastic equilibrium initial condition (\ref{Eq:QuadraticInitialConditions}) by the finite-\(N\) initial condition in Eq.~(\ref{Eq:QuadraticInitialConditionsN}), we find
\begin{align}
\frac{\Omega_S h^2}{0.12}
&\simeq
1.6\times10^{16}
\left(\frac{m_S}{100~{\rm GeV}}\right)^{1/2}
\left(\frac{H_{\rm end}}{7.4\times10^{12}~{\rm GeV}}\right)^2
\nonumber\\
&\times
\left(\frac{N}{10^{10}}\right)
\left(\frac{106.75}{g_*(T_{\rm osc})}\right)^{1/4}
\, .
\label{Eq:OmegaFiniteN}
\end{align}
Comparing this result with Eq.~(\ref{Eq:OmegaQuadraticb}), we see although a finite duration of inflation reduces 
the relic abundance, the improvement is not sufficient to make weak-scale spectator dark matter viable if the oscillation onset occurs during radiation domination. 
Thus, imposing $N\gtrsim N_{\rm iso}$ pushes the
required spectator mass down to $m_S\lesssim3\times10^{-22}$~eV for
our benchmark inflationary scale.

Interestingly, the dependence on the spectator mass is reversed 
relative to that obtained in the stochastic-equilibrium regime. For finite $N$, the random walk variance in Eq.~(\ref{Eq:S2endFiniteN}) is essentially independent of $m_S$, so that the 
spectator energy density at the onset of oscillations scales as $\rho_S(T_{\rm osc})\propto m_S^2$. Since $T_{\rm osc}\propto m_S^{1/2}$ and the subsequent 
matter-like dilution contributes a factor $T_{\rm osc}^{-3}\propto m_S^{-3/2}$, 
the final abundance scales as $\Omega_Sh^2\propto m_S^{1/2}$. Consequently, 
decreasing the spectator mass decreases the relic abundance. This is in contrast 
with the stochastic equilibrium result of Eq.~(\ref{Eq:OmegaQuadraticb}), where $\langle S^2\rangle_{\rm end}\propto m_S^{-2}$ makes the initial spectator energy 
density $\rho_S(T_{\rm osc})$ independent of $m_S$. In that case, the mass 
dependence is controlled entirely by the onset of oscillations: lighter 
spectators begin oscillating later, undergo less matter-like dilution, and therefore produce a larger relic abundance.

\subsubsection{Quartic domination}

In
general, the stochastic dynamics are governed by the full spectator
potential, whose curvature is
\beq
m_{S,{\rm eff}}^2
\simeq
m_S^2
+
3\lambda_S\langle S^2\rangle
\, .
\eeq
\noindent
Once the quartic contribution becomes important
during inflation, both the stochastic distribution and the equilibration time are modified; in particular, in the quartic dominated
regime one finds parametrically $N_{\rm eq}\propto
1/\sqrt{\lambda_S}$. 
Eq.~(\ref{Nisoquad})
then takes the form 
\beq
N_{\rm iso}
\lesssim N
\ll N_{\rm eq}^{(4)}
\sim \frac{4\pi^2(0.132)}{\sqrt{\lambda_S}}
\simeq
\frac{5.2}{\sqrt{\lambda_S}}
\, .
\eeq
which now implies
\begin{align}
\lambda_S
\lesssim
\left(
\frac{5.2}{N_{\rm iso}}
\right)^2
\simeq
2.7\times10^{-19}
\left(
\frac{10^{10}}{N_{\rm iso}}
\right)^2
\, .
\label{Eq:lambdaiso}
\end{align}
This is the quartic analogue of Eq.~(\ref{Eq:miso}), although it
should be understood as a consistency condition for remaining out of
stochastic equilibrium rather than as a direct isocurvature bound on
$\lambda_S$.

 When the quartic term dominates the spectator
potential at the end of inflation, the corresponding initial energy density
is
\beq
\rho_S^{\rm end}
\simeq
\frac{\lambda_S}{4}S_{\rm end}^4
\simeq
\frac{\lambda_S N^2}{64\pi^4}H_{\rm end}^4
\, ,
\label{Eq:QuarticInitialConditionsN}
\eeq
where we have used Eq.~(\ref{Eq:S2endFiniteN})  which in this case assumes $N\ll N_{\rm eq}^{(4)}$. In the quartic regime, the initial field amplitude determines both the
onset of oscillations \emph{and} the subsequent quartic-to-quadratic transition. Thus, the relic abundance must be rederived using this finite $N$
initial condition.

For $a_{\rm osc}<a_m<a_{\rm RH}$, we must first ensure that $a_{\rm osc}<a_{\rm m}$ implying that $S_{\rm end} > \sqrt{2} m_S/\sqrt{ \lambda_S}$. 
Then until the bare mass term dominates $\rho_S \propto S^4 \propto a^{-4}$. Oscillations of $S$ begin when $m_{S,{\rm eff}} \simeq \frac32 H$, where the expansion rate is dominated by the inflaton during reheating. Oscillations begin when
\beq
\frac{a_{\rm osc}}{\aend} = \left( \frac{3 \pi^2}{\lambda_S N} \right)^\frac13.
\eeq
The quadratic term eventually begins to dominate when $S^2 \simeq 2 m_S^2/\lambda_S$ which determines $a_{\rm m}$
\beq
\frac{S_i}{S(a_{\rm m})} = \frac{a_{\rm m}}{a_{\rm osc}} = \frac{\sqrt{\lambda_S N} \Hend}{2 \sqrt{2} \pi m_S}
\eeq
so that \footnote{Note that at $a=a_m$ the density is actually twice that shown since there are equal contributions from quadratic and quartic terms.}   
\beq
\rho_S(a_{\rm m}) = \frac{m_S^4}{ \lambda_S} \, .
\eeq
Then at reheating, we have 
\begin{align}
\rho_S(\arh) = & \rho_S(a_{\rm m}) \left(\frac{a_{\rm m}}{\arh}\right)^3 \nonumber \\ = & \frac{m_S \Hend \sqrt{N} \alpha_{\rm RH} \Trh^4}{16 \sqrt{2} \pi \sqrt{\lambda_S} M_P^2} 
\end{align}
leading to the current relic density given by 
\begin{align}
\frac{\Omega_S h^2}{0.12}
\simeq&
\left(\frac{10^{-20}}{\lambda_S}\right)^{\frac12}
\left(\frac{m_S}{1~{\rm MeV}}\right)
\left(\frac{T_{\rm RH}}{33~{\rm TeV}}\right)
\nonumber\\
&\times
\left(\frac{H_{\rm end}}{7.4\times10^{12}~{\rm GeV}}\right)
\left(\frac{N}{10^{10}}\right)^{\frac12}
\, .
\label{Eq:k2p4N}
\end{align}
Thus, even after imposing the large value of $N$ required by the
isocurvature constraint, the observed relic abundance can still be
obtained in the quartic dominated finite $N$ regime for sufficiently
small self-couplings. For $N\sim10^{10}$, consistency with the
finite $N$ treatment requires approximately
$\lambda_S\ll{\cal O}(10^{-19})$, as discussed above. We see that the correct relic density can be obtained
for a wide range of (correlated) masses, couplings, and
reheating temperatures.
This result also holds for $a_{\rm osc}<a_{\rm RH}<a_m$.

For sufficiently large reheating temperatures such that
$a_{\rm RH}<a_{\rm osc}<a_m$, oscillations begin during radiation domination and 
\beq
\left( \frac{\arh}{a_{\rm osc}} \right)^2 = \frac{3 \sqrt{\lambda_S N} \Hend M_P}{4 \pi \sqrt{\alpha_{\rm RH}} \Trh^2} \, .
\eeq
In this case, 
\beq
\rho_S(a_0) = \frac{\sqrt{2}m_S N^\frac34 \Hend^\frac32 \alpha_{\rm RH}^\frac34}{3^\frac32 4 \lambda_S^\frac14 \pi^\frac32 M_P^\frac32} T_0^3 \frac{g_{*,s}(T_0)}{g_{*}(T_{\rm RH})}\, ,
\eeq
where $T_0$ is the present temperature of the CMB.
This leads to 
\begin{align}
\frac{\Omega_S h^2}{0.12}
\simeq&
\left(\frac{10^{-20}}{\lambda_S}\right)^{\frac14}
\left(\frac{m_S}{6.9~{\rm meV}}\right)
\left(\frac{H_{\rm end}}{7.4\times10^{12}~{\rm GeV}}\right)^{\frac32}
\nonumber\\
&\times
\left(\frac{N}{10^{10}}\right)^{\frac34}
\, .
\label{k4p4N}
\end{align}
\noindent
Once the isocurvature-motivated value $N\sim10^{10}$ is imposed,
together with the requirement that the spectator remain in the
finite $N$ quartic regime, the mass required to reproduce the observed
relic abundance is therefore shifted from the GeV scale to the
meV range.

\begin{figure*}[!ht]
    \centering
    \includegraphics[
        width=\textwidth
    ]{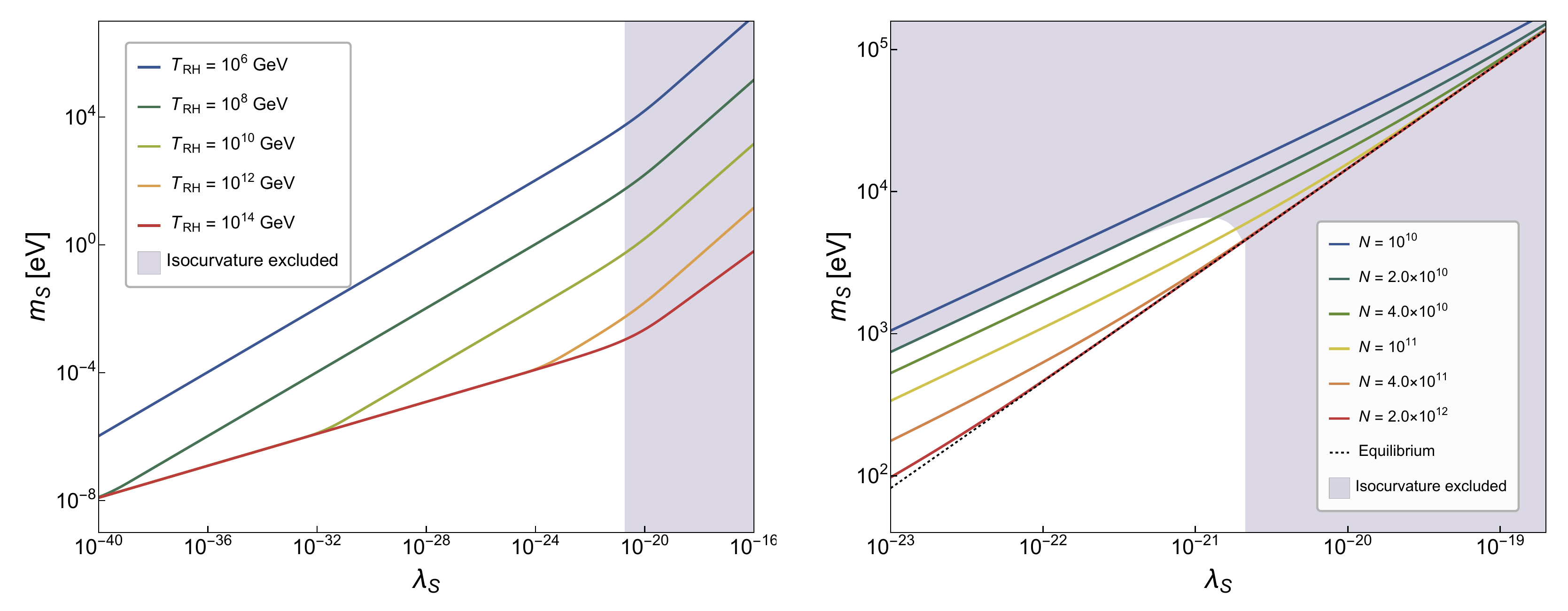}
    \caption{Contours in the $(\lambda_{S},m_S)$ plane compatible with the observed dark matter relic abundance, $\Omega_Sh^2=0.12$, for an isolated self-interacting spectator with no Higgs portal coupling. The left panel depicts contours for several choices of the reheating temperature, $T_{\rm RH}$, with fixed inflationary e-fold number $N=10^{11}$. The right panel depicts contours for several choices of $N$ for fixed $T_{\rm RH}=10^6$~GeV. The purple shaded regions are excluded by isocurvature constraints in Eqs.~(\ref{Eq:IsocurvatureN}) and (\ref{Eq:IsocurvatureQuartic}). The black dotted line in the right panel corresponds to the spectator reaching its equilibrium distribution during inflation.}
    \label{Fig:RelicContoursQuartic}
\end{figure*}

Contours for the relic abundance in the ($\lambda_S, m_S$) plane corresponding to 
$\Trh < \Tosc$ (from Eq.~(\ref{Eq:k2p4N}))
and $\Trh > \Tosc$ (from Eq.~(\ref{k4p4N})) are shown in Fig.~\ref{Fig:RelicContoursQuartic}. The left panel shows the contours for several values of $\Trh$ for $N=10^{11}$ and the right panel shows the contours for several values of $N$ for $\Trh = 10^6$~GeV. For $\Trh < \Tosc$, we see that the resulting relic abundance depends on $\Trh$, and the contours increase as $m_S\propto \sqrt{\lambda_S}$. For fixed $\lambda_S$, smaller values of $m_S$ are required when either $\Trh$ or $N$ are increased. 
For $\Trh > \Tosc$, $m_S \propto \lambda_S^\frac14$ independent of $\Trh$ which explains why the curves merge at low $\lambda_S$ and $m_S$ as seen in the left panel. This separation occurs when
\beq
\lambda_S \simeq \frac{64\pi^2 \Trh^4}{27 \beta N \Hend^2 M_P^2} \, ,
\eeq
when $\Trh \simeq \Tosc$  and we have defined $\beta\equiv\frac{40}{\pi^2 g_*(T_{\rm osc})}$.

In the right panel of Fig.~\ref{Fig:RelicContoursQuartic}, the area shown always corresponds to $\Trh < \Tosc$. Thus the contours in the finite $N$ regime again have the slope $m_S \propto \sqrt{\lambda_S}$. For a given value of $\lambda_S$, the equilibrium value of $N$ is $4 \pi^2 (.132)/\sqrt{\lambda_S}$. Therefore for each value of $N$, as $\lambda_S$ is increased, the equilibrium value of $N$ decreases and results in the merging of the contours to a common value of $N(\lambda_S)$. From Eq.~(\ref{Eq:k2p4}), we see that along this common contour, $m_S \propto \lambda_S^\frac34$, which is depicted as a black dotted line.

Now,  as we will see in the remainder 
of the work, the introduction of a feeble Higgs portal interaction can 
drastically modify these results.



\section{Higgs portal and the hierarchy of the effective masses}
\label{Sec:EffectiveMassHierarchy}

\subsection{Effective spectator mass hierarchy determines the oscillation sequence}
\label{OscHistory}

The hierarchy of the effective mass terms in Eq.~(\ref{Eq:EffectiveMassIntro}), will drive nearly all of the important physics in this work, including the sequence of oscillations of the spectator condensate and the stochastic inflationary dynamics. Because there are multiple possible scenarios for the manner in which the spectator condensate may undergo oscillations after inflation, it will be helpful to establish the basic conditions under which each of the effective mass terms will drive oscillations. Because we will generally treat reheating as a non-instantaneous process, each term will then have the possibility of driving spectator oscillations either during the reheating period or during radiation domination. In this section, we first provide analytic expressions for the oscillation temperatures ($T_{\rm osc}$) associated with each of these possibilities. With three effective mass terms \footnote{Note that although there are actually four effective mass terms shown in Eq.~(\ref{Eq:EffectiveMassIntro}), the last term which is proportional to the Higgs VEV after EWSB will not be responsible for oscillation onset in our parameter space of interest.} and two cosmological eras under consideration, we find six distinct possible oscillation onset temperatures, which we summarize in Table \ref{Table1} below. In each case, we use the condition
given in Eq.~(\ref{Eq:QuadraticOscillation}) to determine the oscillation temperature.

\begin{table*}[ht]
    \centering
    \renewcommand{\arraystretch}{2.2}
    \caption{Oscillation temperatures for the three contributions to the effective scalar mass during radiation domination and non-instantaneous reheating. We have defined $\beta\equiv\frac{40}{\pi^2 g_*(T_{\rm osc})}$ and we have taken $\kappa=1$.}
    \label{tab:oscillation_temperatures_combined}

    \begin{tabular*}{\textwidth}{@{\extracolsep{\fill}} l c c}
        \hline\hline
        \multicolumn{1}{c}{Effective mass term}
        &
        Radiation domination
        &
        Non-instantaneous reheating
        \\
        \hline

        Bare mass,
        $m_{\rm eff}^2 \supset m_S^2$
        &
        $\displaystyle
        T_{\rm osc}^{(m)}
        = (\frac{9}{16}\beta)^{1/4}\sqrt{m_SM_P}
        $
        &
        $\displaystyle
        T_{\rm osc}^{(m)}
        = \beta^{1/8}
        (T_{\rm RH}^2m_SM_P)^{1/4}
        $
        \\

        Thermal mass,
        $m_{\rm eff}^2 \supset c_S T^2$
        &
        $\displaystyle
        T_{\rm osc}^{(\rm th)}
        =
        \frac{3}{4}\sqrt{\beta c_S}\,M_P
        $
        &
        $\displaystyle
        T_{\rm osc}^{(\rm th)}
        =
        (\beta c_S)^{1/6}
        (T_{\rm RH}^2M_P)^{1/3}
        $
        \\

        Radiative quartic,
        $m_{\rm eff}^2 \supset 3\lambda_S^{\rm rad}S^2$
        &
        $\displaystyle
        T_{\rm osc}^{(\rm rad)}
        =
        \sqrt{\frac{3(0.16)}{16\pi}} \,\beta^{1/4}
        \lambda_{HS}^{1/4}
        \rho_{\rm end}^{1/4}
        $
        &
        $\displaystyle
        T_{\rm osc}^{(\rm rad)}
        =
        \left(\frac{0.16}{4\pi}\right)^{1/4} \,\beta^{1/8}
        \lambda_{HS}^{1/8}
        (\sqrt{\rho_{\rm end}}T_{\rm RH}^2)^{1/4}
        $
        \\

        \hline\hline
    \end{tabular*}

    \begin{minipage}{\textwidth}
        \footnotesize
        \raggedright
        \textit{Note:} The expressions shown
    here assume that the
    spectator remains approximately frozen, $S\simeq S_{\rm end}$,
    until the onset of oscillations. For the radiative quartic term, unlike the bare and thermal mass terms, $T_{\rm osc}$ depends on the initial conditions for $S$. Recall that the stochastic equilibrium variance for $S$ is obtained using the joint two-field evolution of the spectator and the Higgs, as described in Appendix \ref{app:joint}. The third row expressions use Eq.~(\ref{Eq:JointEvolutionS}) for the spectator initial conditions. 
    \end{minipage}
\label{Table1}
\end{table*}

Alongside the analytic expressions for the oscillation temperatures themselves, we must also enforce consistency conditions to ensure that the respective oscillation temperatures fall within their domains of validity. In particular, for oscillations beginning during the reheating era, we require $T_{\rm RH}<T_{\rm osc}<T_{\rm max}$, while for oscillations during radiation domination we simply require $T_{\rm osc}<T_{\rm RH}$. The reheating era conditions can be written as follows: \footnote{Of course, the upper bound on $\lambda_{HS}$ for the radiative quartic is not phenomenologically significant since perturbativity is a more stringent constraint.}
\begin{align}
\text{Bare mass:}\qquad
&\frac{T_{\rm RH}^2}{M_P}\beta^{-1/2}
< m_S <
\frac{5}{2}\left(\frac{3}{8}\right)^{8/5}H_{\rm end},
\nonumber \\
\text{Thermal mass:}\qquad
&6\beta^{-1}\frac{T_{\rm RH}^2}{M_P^2}
<
\lambda_{HS}
< 0.96 \beta^{-1/4}\,
\frac{\rho_{\rm end}^{3/4}}{T_{\rm RH}M_P^2},\nonumber \\
\text{Rad. quartic:}\qquad
&6.2\times10^3\,\frac{T_{\rm RH}^4}{\rho_{\rm end}}\beta^{-1}
<
\lambda_{HS}
<
550.
\end{align}

It is then useful to distinguish four possible physical sequences for the spectator condensate's oscillations. The following four sequences are possible: 
\begin{enumerate}
\item Bare mass oscillations alone
\item Quartic oscillations $\rightarrow$ bare mass oscillations
\item Thermal oscillations $\rightarrow$ bare mass oscillations
\item Quartic oscillations $\rightarrow$ thermal oscillations $\rightarrow$ bare mass oscillations
\end{enumerate}

\noindent
It is not difficult to see why these are the only possible allowed sequences 
from the three contributions to the effective mass. First, after bare mass 
oscillations have begun, the other two contributions to the effective mass 
cannot overtake the bare mass, since the latter is constant while the former 
two are decreasing with cosmological time. Second, it is also not possible 
for quartic oscillations to follow thermal oscillations. This can be seen from the following argument.
During reheating, we have $T\propto a^{-3/8}$ and we will see that after the 
onset of thermal oscillations, $S \propto a^{-21/16}$. Because the thermal 
and quartic contributions to the effective mass are proportional to $T$ and $S$ respectively, we see that once the thermal term is dominant, the quartic will never exceed it. 
By contrast, it is in fact possible for quartic oscillations to begin during 
reheating followed by the onset of thermal oscillations. Similarly, for high 
reheating temperature, oscillations begin 
during radiation domination where $T\propto a^{-1}$, and in this case we 
shall see that after thermal oscillations begin we have $S\propto a^{-1}$. 
Therefore both contributions to the effective mass scale as $m_{S,\rm eff}\propto a^{-1}$ such that the quartic can never overtake the thermal term once the latter has initiated oscillations. Therefore, we are left 
with the four possible sequences of oscillations enumerated above. 

The first two sequences mentioned above are possible without the Higgs portal, and we have provided 
the key results for these scenarios in the preceding section. In contrast, 
sequences 3 and 4 are not possible for an isolated spectator decoupled from the thermal bath, but become possible in the presence of the Higgs portal 
interaction. Hence, scenarios 3 and 4 are the main focus of this work. Scenario 2 is also of interest in the setting of the Higgs portal, since it is possible for the Higgs portal to induce oscillations through the radiative quartic term followed by bare mass oscillations without an intervening thermal regime.  For the reader's convenience, we also include Table \ref{Table2} which contains analytic expressions for the 
energy density of the condensate at the onset of oscillations induced by 
each of the three terms in the effective mass. These will be used repeatedly 
throughout the following analysis.

\begin{table*}[!ht]
    \centering
    \renewcommand{\arraystretch}{2.3}
    \caption{Analytic expressions for the energy density of the spectator condensate at the onset of oscillations in different scenarios. We include results for when the first onset of oscillations is induced by the bare mass, thermal mass, or self-interaction terms for both finite $N$ and stochastic equilibrium initial conditions. The bare and self-interaction results hold for both radiation domination (RD) and reheating (Reh) while the thermal oscillation results differ in the two epochs and are distinguished accordingly.}

    \begin{tabular*}{\textwidth}{@{\extracolsep{\fill}} l c c}
        \hline\hline
        \multicolumn{1}{c}{Effective mass term}
        &
        Finite $N$
        &
        Stochastic equilibrium
        \\
        \hline

        Bare mass,
        $m_{\rm eff}^2 \supset m_S^2$
        &
        $\displaystyle
        \rho_S (T_{\rm osc}^{(m)})
        = \frac{N}{24 \pi^2}\rho_{\rm end}\left(\frac{m_S}{M_P}\right)^2
        $
        &
        $\displaystyle
        \rho_S(T_{\rm osc}^{(m)})
        = \frac{1}{48 \pi^2} \frac{\rho_{\rm end}^2}{M_P^4}
        $
        \\

        Radiative quartic,
        $m_{\rm eff}^2 \supset 3\lambda_S^{\rm rad}S^2$
        &
        $\displaystyle
        \rho_S(T_{\rm osc}^{(\rm rad)})
        =
        \frac{N^2 \lambda_{HS}^2}{9216 \pi^6}\frac{\rho_{\rm end}^2}{M_P^4}
        $
        &
        $\displaystyle
        \rho_S(T_{\rm osc}^{(\rm rad)})
        =
        \frac{(0.16)^4}{576\pi^2}\frac{\rho_{\rm end}^2}{M_P^4}
        $
        \\

        Thermal mass,
        $m_{\rm eff}^2 \supset c_S T^2$ (RD)
        &
        $\displaystyle
        \rho_S(T_{\rm osc}^{(\rm th)})
        =
        \frac{3 N \beta}{128 \pi^2}(c_S )^2\rho_{\rm end}
        $
        &
        $\displaystyle
        \rho_S(T_{\rm osc}^{(\rm th)})
        =
        \frac{(0.05)\beta c_S}{64}\rho_{\rm end}
        $
        \\

        Thermal mass,
        $m_{\rm eff}^2 \supset c_S T^2$ (Reh)
        &
        $\displaystyle
        \rho_S(T_{\rm osc}^{(\rm th)})
        =
        \frac{N \beta^{1/3}(c_S )^{4/3}}{24 \pi^2}\frac{T_{\rm RH}^{4/3}\rho_{\rm end}}{M_P^{4/3}}
        $
        &
        $\displaystyle
        \rho_S(T_{\rm osc}^{(\rm th)})
        \!=\!
        \frac{(0.05)}{36}\!(\beta c_S)^{1/3}\!\left(\!\frac{T_{\rm RH}}{M_P}\!\right)^{4/3}\!\rho_{\rm end}$
        \\

        \hline\hline
    \end{tabular*}

    \begin{minipage}{\textwidth}
        \footnotesize
        \raggedright
        \textit{Note:} For stochastic equilibrium expressions in the right-hand column, we have assumed quadratic-dominated equilibrium for the bare mass entry, while for the thermal and radiative quartic entries we have used Eq.~(\ref{Eq:JointEvolutionS}). We have defined $\beta\equiv\frac{40}{\pi^2 g_*(T_{\rm osc})}$, and we have taken $\kappa=1$. 
       We have also assumed that the bare quartic coupling is negligible so that the effective quartic coupling is dominated by the radiative contribution.  
    \end{minipage}
\label{Table2}
\end{table*}

\subsection{Effective mass hierarchy and the stochastic inflationary evolution}

In this work, we consider both the case where inflation is sufficiently long for the spectator to reach its equilibrium variance of fluctuations, as well as the case where the number of e-folds $N$ is insufficient to reach equilibrium. In the latter case of shorter inflation, the stochastic evolution of the spectator is akin to a random walk when $N \ll H^2/m_{S,\rm eff}^2$. As a result, the final amplitude of the fluctuations will be independent of the effective mass hierarchy, and will instead depend only on $N$, as shown in  Eq.~(\ref{Eq:S2endFiniteN}). In this case, the subsequent oscillation history will be sensitive to the effective mass hierarchy, but the initial conditions for the fluctuations coming out of inflation will not.

For the case in which the spectator condensate reaches its equilibrium 
variance during inflation (large $N$), the initial conditions for the fluctuations at the end of inflation \textit{and} the oscillation history will both be determined by the Higgs portal interaction in our parameter space of interest. In particular, we cannot simply recycle the results for the equilibrium variance for an isolated quadratic or quartic spectator, because our Higgs portal interaction will generally modify these results. Instead, we must employ a two-field, joint stochastic evolution of both $S$ and $h$ using $V(S,h)$, rather than treating $S$ as an isolated field. We supply the details of this approach in Appendix \ref{app:joint}.

The result of performing the joint evolution is captured in Eqs.~(\ref{Eq:JointEvolutionS}) and (\ref{Eq:JointEvolutionS2}) where either $\langle |S|\rangle_{\rm eq,joint}$ or $\langle S^2\rangle_{\rm eq, joint}$ will supply our initial condition for the post-inflationary evolution. \footnote{Whether $\langle |S|\rangle_{\rm eq,joint}$ or $\langle S^2\rangle_{\rm eq, joint}$ is needed for the initial condition is determined by whether the final relic abundance is sensitive to $|S|_{\rm end}$ or $S_{\rm end}^2$. For the quartic $\rightarrow$ thermal $\rightarrow$ bare sequence we will need the former, while for the  thermal $\rightarrow$ bare sequence we will need the latter. This distinction affects the calculated abundance by a simple $\mathcal{O}(1)$ factor} This results in 
\beq
N_{\rm  eq} \simeq \frac{(0.51)^2 \sqrt{\lambda_H} 4 \pi^2}{\lambda_{HS}} \simeq \frac{1.0}{\lambda_{HS}} \left(\frac{\lambda_H}{0.01}\right)^{1/2}\, .
\label{jointN}
\eeq 
In our analytical and numerical results, we take $\lambda_H=0.01$ for specificity, and it is straightforward to generalize to other choices. This joint evolution treatment leads to spectator fluctuations at the end of inflation which are suppressed by approximately a factor of 
30 in $\langle S^2 \rangle_{\rm end}$ or by a factor of about 7 in the 
characteristic amplitude $\langle |S|\rangle_{\rm end}$ relative to the equilibrium fluctuations supplied by an isolated radiative quartic term. Eq.~(\ref{jointN}) above uses $\langle |S|\rangle_{\rm end}$ for the initial condition.

We can obtain some intuition for Eqs.~(\ref{Eq:JointEvolutionS}) and (\ref{jointN}) in the following manner. We recall that during inflation, the effective spectator mass takes the form $m_{S,\rm eff}^2 = m_S^2+3\lambda_S^{\rm rad}S^2+\lambda_{HS}h^2/2$ where $h$ is the $\mathcal{O}(4)$ radial Higgs field in the symmetric phase (note that there will not be a thermal effective mass term during the cold inflation epochs we consider). Naively, one might anticipate that the stochastic equilibrium distribution could be dominated by any of the three effective mass terms for the spectator above. However, in practice, we find that the $\lambda_{HS} h^2/2$ term is dominant over essentially our entire parameter space of interest. For $\lambda_H=0.01$, we find that $\langle h^2 \rangle\sim\mathcal{O}(3) H^2$ for all of our parameter space, such that the final amplitude of the Higgs fluctuations is nearly independent of the portal interaction for the feeble $\lambda_{HS}$ we consider. If we use Eq.~(\ref{Sisat}) with $m_{S,\rm eff}^2=\lambda_{HS}\langle h^2\rangle/2$, we recover parametrically the result shown in Eq.~(\ref{Eq:JointEvolutionS}). The full numerical result is different by about a factor of 2. In summary, in our parameter space of interest the joint stochastic evolution of the spectator will not be dominated by either the bare mass term nor the radiative quartic term, but always the $\lambda_{HS}h^2S^2$ term. 

Having established the basic manner in which the effective spectator mass governs both the inflationary dynamics and the post-inflationary evolution, we now proceed to our results section where we derive the final abundance in each relevant regime and report our full numerical results and the associated parameter space. 




\section{Thermal effects on the evolution of the condensate}
\label{Sec:ThermalReshaping}

Of the four possible histories we outlined in Section \ref{OscHistory}, we have already reviewed the results of the first two for an isolated spectator in Section \ref{Sec:InflationaryMemory}. 
Namely, we provided the relic density when the bare mass term always 
dominates (Eq.~(\ref{Eq:casek2}) for $T_{\rm osc} > \Trh$, Eq.~(\ref{Eq:OmegaQuadraticb}) for $T_{\rm osc} < \Trh$  in stochastic 
equilibrium, and Eq.~(\ref{Eq:FiniteNreh}) and Eq.~(\ref{Eq:OmegaFiniteN}) 
for the same two cases when $N$ is smaller than its equilibrium value). We also provided the relic density when quartic oscillations dominate initially and eventually yield to quadratic (bare mass) 
oscillations (scenario 2). This is given in Eqs.~(\ref{Eq:k2p4}) and (\ref{k4p4}) for 
the two orderings of $\Tosc$ and $\Trh$ with maximal $N$ and in 
Eqs.~(\ref{Eq:k2p4N}) and (\ref{k4p4N}) for small $N$. Both of these cases 
are possible without the Higgs portal coupling.  

We now turn to the novel spectator histories involving a very feeble Higgs portal,
corresponding to scenarios 3 and 4 of Section~\ref{OscHistory}. In these cases, after inflation 
the spectator condensate will acquire a thermal mass from its interactions 
with the Higgs boson in the thermal bath which may qualitatively change 
the condensate evolution. We will assume through our remaining analysis that the bare quartic coupling is sufficiently small that it is negligible. 
However, we do keep in mind the  radiative contribution to $\lambda_S$ induced from the coupling to the Higgs boson, $\lambda_{HS}$, such that we will have $\lambda_{S}=\lambda_{S}^{\rm rad}$. We consider both stochastic equilibrium and finite $N$ initial conditions
for the condensate at the end of inflation, and subsequently determine
which of these possibilities remain compatible with isocurvature and
dark matter abundance constraints. 


\subsection{Thermal $\rightarrow$ bare mass oscillations}
\label{Sec:thermalToBare}

In this subsection, we consider the case that the effective mass is dominated by the thermal contribution at the time oscillations begin. Assuming instantaneous thermalization at $\Tmax$,
we therefore minimally require that
$c_S \Tmax^2>m_S^2$. 
Oscillations begin when Eq.~(\ref{Eq:QuadraticOscillation}) is satisfied with $m_{S,{\rm eff}} \simeq \sqrt{c_S} T$.  Depending on $\Trh$, oscillations may begin before or after reheating. For each of the two cases, we obtain, 
\begin{align}
T_{\rm osc}^6
=
\beta c_S T_{\rm RH}^4M_P^2
\, ,
\qquad
T_{\rm osc}>T_{\rm RH}
\, \\
T_{\rm osc}^2
=
\frac{9}{16} \beta c_S\,M_P^2
\, ,
\qquad
T_{\rm osc}<T_{\rm RH}
\, ,
\end{align}
as given in the middle row of Table \ref{tab:oscillation_temperatures_combined}.

These determinations assume the stronger condition
$c_S T_{\rm osc}^2>m_S^2$, which gives
\begin{align}
m_S &\lesssim
\beta^{1/6}c_S^{2/3}
\left(T_{\rm RH}^2M_P\right)^{1/3}
\, ,
\qquad T_{\rm osc}>T_{\rm RH}
\, \label{Eq:TOscCondition1}
\\
m_S &\lesssim
\frac{3}{4}\sqrt{\beta}\,c_S M_P
\, ,
\qquad T_{\rm osc}<T_{\rm RH}
\, .
\label{Eq:ToscConditions}
\end{align}
\noindent
These conditions demarcate the boundaries between the ``bare mass oscillation only" scenario and the ``thermal oscillation $\rightarrow$ bare mass oscillation" scenario (corresponding to sequences 1 and 3 described in Section \ref{Sec:EffectiveMassHierarchy}), in the case where the radiative quartic contribution is comparatively negligible. 

As we will see, during reheating, $m_{S,{\rm eff}}/H\propto a^{\frac98} \propto T^{-3}$ and during
radiation domination,
$m_{S,{\rm eff}}/H\propto a \propto T^{-1}$. In both cases this ratio  {\it increases} as the
Universe cools. The thermal Higgs contribution can consequently trigger
spectator oscillations much earlier than the vacuum mass alone.

In this subsection, we consider the case where the thermal contribution to the effective mass exceeds the quartic term's contribution at the onset of oscillations (we consider the alternative in a later subsection). As discussed above, the  radiatively induced quartic coupling grows as $\lambda_{HS}^2$. Therefore, requiring the thermal contribution to dominate over the quartic one at the onset of oscillations during reheating imposes an upper limit on $\lambda_{HS}$. This condition can be written as
\beq
c_S T^2
=
\frac{1}{6}\lambda_{HS}T^2
>
\frac{3\kappa}{16\pi^2}\lambda_{HS}^2 S_{\rm end}^2
\, .
\label{thermmassbig}
\eeq
\noindent
Evaluating this condition at $T_{\rm osc}$, we find for
$T_{\rm osc}>T_{\rm RH}$
\begin{align}
\lambda_{HS}
&<
\frac{128\pi^6}{27\sqrt{3}}\,
\beta^{1/2}
\frac{T_{\rm RH}^2M_P}
{\kappa^{3/2}N^{3/2}H_{\rm end}^3}
\nonumber\\
&\simeq
3.1\times10^{-17}
\left(\frac{T_{\rm RH}}{10^8~{\rm GeV}}\right)^2
\left(\frac{10^{10}}{N}\right)^{3/2}
\nonumber \\
& \times
\left(\frac{7.4\times10^{12}~{\rm GeV}}
{H_{\rm end}}\right)^3
\, ,
\label{ulimitlhs}
\end{align}
\noindent
where we have taken $\kappa=1$ in the final equality.

We are of course also assuming here that $c_S T_{\rm osc}^2>m_S^2$, which gives a lower limit on
$\lambda_{HS}$ which can be found directly from Eq.~(\ref{Eq:TOscCondition1}).
Combining Eqs.~(\ref{Eq:TOscCondition1}) and (\ref{ulimitlhs}), we obtain an
upper limit on $m_S$ for fixed $N$ and $H_{\rm end}$. For our standard
$H_{\rm end}=7.4\times10^{12}~{\rm GeV}$, this gives
\beq
m_S
<
0.50~{\rm GeV}
\left(\frac{10^{10}}{N}\right)
\left(\frac{T_{\rm RH}}{10^8~{\rm GeV}}\right)^2
\, .
\eeq

As $N$ increases, the upper limit in Eq.~(\ref{ulimitlhs}) decreases,
while the lower limit on $\lambda_{HS}$ remains unchanged.
The two conditions can therefore be simultaneously satisfied only for
\beq
N<N_{\rm max}
=
\frac{16\pi^4\beta^{\frac12}}{27\kappa}
\frac{T_{\rm RH}^2M_P}{H_{\rm end}^2m_S}
\, .
\eeq
For $N > N_{\rm max}$ a dominant thermal mass at the time of oscillations is not possible for $T_{\rm osc}>T_{\rm RH}$. 
$N_{\rm max}$ is equal to the (quadratic) equilibrium value
when 
\beq
m_S = \frac{81}{32\pi^4} \frac{\kappa}{\sqrt{\beta}} \frac{\Hend^4}{\Trh^2 M_P} \, ,
\label{msnmax}
\eeq
and the portal coupling is
\begin{align}
& \lambda_{HS}  = \frac{2187 \kappa^\frac32 \Hend^6}{64\sqrt{2}\pi^6 \beta \Trh^4 M_P^2} \nonumber \\
 & = 1.8\times 10^{-8} \left(
\frac{H_{\rm end}}{7.4\times10^{12}~{\rm GeV}}
\right)^6 \left( \frac{10^{12}~{\rm GeV}}{\Trh} \right)^4 \, .
\label{lhsnmax}
\end{align}

For the case $T_{\rm osc}<T_{\rm RH}$, the dependence on
$\lambda_{HS}$ cancels between the two sides of
Eq.~(\ref{thermmassbig}). Consequently, this condition does not impose
a bound on $\lambda_{HS}$, but instead yields an upper bound on $N$
for the thermal contribution to dominate at the onset of oscillations,
\beq
N<N_{\rm max}
=
\frac{\pi^4}{3\kappa}\,
\beta
\left(\frac{M_P}{H_{\rm end}}\right)^2
\simeq 1.3\times10^{11}
\, ,
\eeq
\noindent
where the numerical value assumes
$H_{\rm end}=7.4\times10^{12}~{\rm GeV}$ and $\kappa=1$.
Note that this remains compatible with the isocurvature requirement
$N\gtrsim  10^{10}$.

For comparison, the equilibrium variance of $S$ fluctuations is determined by Eq.~(\ref{Eq:JointEvolutionS}).
Requiring $N_{\rm eq}<N_{\rm max}$ therefore gives
\beq
\lambda_{HS} \gtrsim \frac{(0.05) 4\pi^2}{N_{\rm max}}=1.5\times10^{-11}.
\eeq
\noindent
As we will see below, this lower bound on $\lambda_{HS}$
prevents the stochastic equilibrium solution for the thermal $\rightarrow$ bare mass oscillation history from self-consistently
reproducing the observed relic abundance. The physical reason is because the radiative quartic term will instead drive the onset of oscillations when this limit on $\lambda_{HS}$ is imposed and the correct abundance is required.

For $T>T_{\rm osc}$, the spectator remains overdamped,
$S\simeq S_{\rm end}$, as long as the quartic contribution remains
subdominant. The subsequent evolution of the energy density depends on
whether the onset of oscillations occurs before or after the completion
of reheating. We consider these two cases in turn.


\subsubsection{$T_{\rm osc}>T_{\rm RH}$}

In this section, we will first derive results for the finite $N$ (shorter inflation) case, after which we will provide the corresponding stochastic equilibrium results. When the thermal contribution dominates the effective mass of the spectator, the energy density at $T = \Tosc$ is
\begin{align}
\rho_S (T_{\rm osc})&=\frac{1}{2}{c_S}  T_{\rm osc}^2 \langle S^2 \rangle_{\rm end} \nonumber \\ &= \frac{N \rho_{\rm end}}{24 \pi^2} {c_S}^{4/3} \beta^{1/3}\left(\frac{T_{\rm RH}}{M_P}\right)^{4/3}\,,
\label{rhoattosclowreh}
\end{align}

\noi 
where we used Eq.~(\ref{Eq:S2endFiniteN}) for
$\langle S^2 \rangle_{\rm end}$.
This corresponds to the Finite $N$ entry in the 4th row of Table \ref{Table2}. 
We can determine the redshifting of the spectator energy density directly from the
adiabatic evolution equation
\begin{align}
\dot{\rho}_S
=
\left(
-3H+\frac{\dot m_{S,\rm eff}}{m_{S,\rm eff}}
\right)\rho_S
\, ,
\label{Eq:mdot}
\end{align}
which is easily derived from the equation of motion for the homogeneous condensate, $S$.
Eq.~(\ref{Eq:mdot}) then immediately implies the existence of the following adiabatic invariant, which governs much of the spectator dynamics in our model
\beq
\frac{\rho_S(a) a^3}{m_{S,\rm eff}(a)}=constant.
\label{Eq:AdiabaticInvariant}
\eeq
\noindent 
When the thermal mass is dominant during matter-like reheating, we have
$m_{S,\rm eff}\propto T\propto a^{-3/8}$.
 Therefore,
$\dot m_{S,\rm eff}/m_{S,\rm eff}=-3H/8$, which gives
\begin{align}
\dot{\rho}_S+\frac{27}{8}H\rho_S=0
\,.
\label{Eq:RhoSRH}
\end{align}
\noindent

Because the thermal mass decreases with time while the bare mass term remains constant, the bare mass will eventually overtake the thermal mass at a temperature $T_\times$ defined by
\beq
c_S T_\times^2 = m_S^2
\label{Eq:Tcross}
\eeq
and we are implicitly assuming that $\Tosc > T_\times$ and $\Tosc > T_{\rm EW}$ 
to ensure that the thermal correction induces the oscillations and that they occur above the 
electroweak scale which guarantees the presence of Higgs bosons in the thermal bath. Eq.~(\ref{Eq:RhoSRH}) then leads to
\begin{align}
\rho_S(a)
\simeq
\rho_S(a_{\rm osc})
\left(\frac{a_{\rm osc}}{a}\right)^{27/8}, \hspace{1mm}
(a_{\rm osc}<a< \text{min}(a_{\times},a_{\rm RH}))
\,
\label{Eq:rhoSduringReheatinga}
\end{align}
or 
equivalently, 
\begin{align}
\rho_S(T)
\simeq
\rho_S(T_{\rm osc})
\left(\frac{T}{\Tosc}\right)^{9}, \hspace{1mm}
(\text{max}(T_\times,\Trh) < T < \Tosc)
\, ,
\label{Eq:rhoSduringReheating}
\end{align} where $a=a_{\times}$ is the scale factor corresponding to $T=T_\times$. This scaling is one of the notable results of our analysis: thermal mass-induced
oscillations during matter-like reheating lead to a characteristic
$\rho_S\propto a^{-27/8}$ dilution of the primordial condensate. This result is valid so long as the thermal correction dominates the effective mass. 

After reheating, while the thermal contribution still dominates, the
Universe is radiation dominated and $m_{S,\rm eff}\propto T\propto a^{-1}$.
The adiabatic evolution (\ref{Eq:mdot}) therefore gives $\rho_S\propto a^{-4}$ until
$T=T_\times$.
Eq.~(\ref{Eq:rhoSduringReheating}) captures the main impact of thermal oscillations during non-instantaneous reheating on the condensate's energy density, which will ultimately determine the final relic abundance.
At temperatures below $T_\times$, the spectator potential is again dominated by its
vacuum quadratic term. The condensate therefore behaves as ordinary
non-relativistic matter and its energy density redshifts as
$\rho_S\propto a^{-3}$.

To calculate the relic abundance in this case, we can evolve the energy density from $\Tosc$ in Eq.~(\ref{rhoattosclowreh})
to the present. 
We can then
follow the condensate from the onset of thermal oscillations through the
transition to bare mass domination, down to the present epoch. For finite
inflation and considering $T_{\rm osc}>T_{\rm RH}>T_\times$ for definiteness, we obtain
\bea
\rho_S(T_0)
&=&
\rho_S(T_{\rm osc})
\left(\frac{\Trh}{\Tosc}\right)^{9}
\left(\frac{T_\times}{\Trh}\right)^4
\left(\frac{T_0}{T_\times}\right)^3
\nonumber\\
&=&
\frac{N\rho_{\rm end}}{24 \pi^2}
c_S^{-2/3} \beta^{-\frac76}
\left(\frac{T_0^3m_S}{M_P^4}\right)
\left(\frac{T_{\rm RH}}{M_P}\right)^{1/3} \nonumber \\
&& \times \frac{g_{*s}(T_0)}
{g_{*s}(T_\times)}
\label{rhosto}
\, .
\eea

Using Eq.~(\ref{rhosto}), the relic abundance can then be written as
\begin{align}
\frac{\Omega_S h^2}{0.12}
&\simeq
\left(\frac{N}{10^{10}}\right)
\left(\frac{H_{\rm end}}
{7.4\times10^{12}~{\rm GeV}}\right)^2
\left(\frac{m_S}{0.46~{\rm eV}}\right)
\nonumber\\
&\quad\times
\left(\frac{10^{-14}}{\lambda_{HS}}\right)^{2/3}
\left(\frac{T_{\rm RH}}{4~{\rm MeV}}\right)^{1/3}
\, .
\label{Eq:Omegah2quadraticNTh}
\end{align} 
We have normalized the reheating temperature to
$T_{\rm RH}=4~{\rm MeV}$, the lower limit allowed by BBN \cite{tr4},
since this maximizes the spectator mass compatible with the observed
relic abundance and therefore provides the most conservative benchmark
for testing the viability of this branch. However, while Eq.~(\ref{Eq:Omegah2quadraticNTh}) correctly conveys the scaling associated with each parameter, our displayed normalization is not internally consistent with the thermal $\rightarrow$ bare history. Rather, for this parameter set, the quartic term will actually induce oscillations prior to the thermal term, yielding a quartic $\rightarrow$ thermal $\rightarrow$ bare history which we will discuss in Section \ref{Sec:QTB}. This is one of the primary limitations associated with scenario 3. In particular, it is difficult to satisfy the correct abundance while simultaneously satisfying isocurvature constraints and thermal-first oscillation onset. We will see in Section \ref{Sec:Nonmin} that considering non-minimal gravity is one way to fundamentally recover the thermal $\rightarrow$ bare history. Furthermore, the normalization shown in Eq.~(\ref{Eq:Omegah2quadraticNTh}) is also inconsistent with a bound on the DM mass which we now turn to.

The isocurvature constraint requires
$N\gtrsim N_{\rm iso}\simeq 10^{10}$. Even taking the reheating
temperature close to its lowest value allowed by BBN, Eq.~(\ref{Eq:Omegah2quadraticNTh})
therefore requires a very small spectator mass. For
$\lambda_{HS}=10^{-13}$, for example, one finds
$m_S\lesssim 2.1~{\rm eV}$.
 However, below the electroweak
scale, this is incompatible with the unavoidable contribution to the physical
spectator mass induced by the Higgs VEV given in Eq.~(\ref{vcontr}),
\beq
m_{S,\rm eff}
\geq
\sqrt{\frac{\lambda_{HS}}{2}}\,v
=
\sqrt{3c_S}\,v
\, ,
\label{Eq:masscondition}
\eeq
when $m_S^2 \ge 0$.
For $\lambda_{HS}=10^{-13}$, this implies
$m_{S,\rm eff}\gtrsim55~{\rm keV}$, more than four orders of magnitude
above the mass required by the relic abundance even for a reheating
temperature close to the BBN limit. This illustrates the strong tension
between the isocurvature constraint and the Higgs VEV contribution in the
quadratic finite-$N$ regime.

The result in Eq.~(\ref{Eq:Omegah2quadraticNTh}) applies to finite $N$ shorter inflation. In the
stochastic equilibrium limit, the initial conditions for the $S$ fluctuations from the joint, two-field stochastic evolution (see Appendix \ref{app:joint}) are achieved for thermal to bare sequence by taking $N\rightarrow(0.05)4\pi^2/\lambda_{HS}$ in Eq.~(\ref{Eq:Omegah2quadraticNTh}), which gives
\begin{align}
\frac{\Omega_S h^2}{0.12}
&\simeq 9.4\times10^{8}
\left(\frac{\Hend}{7.4\times10^{12}~{\rm GeV}}\right)^2
\left(\frac{m_S}{10^7~{\rm GeV}}\right)
\nonumber\\
&\times
\left(\frac{10^{-7}}{\lambda_{HS}}\right)^{5/3}
\left(\frac{T_{\rm RH}}{4~{\rm MeV}}\right)^{1/3}
\, .
\label{Eq:QuarticRehRelic}
\end{align}
where again we find the spectator to be hopelessly overproduced.



\subsubsection{$T_{\rm osc}<T_{\rm RH}$}

In this case, the Universe becomes radiation dominated before the spectator starts to oscillate. However, the energy density, which is dominated by the thermal mass contribution, already begins to evolve before the oscillations begin.
Prior to radiation domination, $T\propto a^{-3/8}$ and the energy density of $S$ evolves as
\beq
\rho_S 
\simeq
\frac{1}{2}c_ST^2S_{\rm end}^2
\propto a^{-\frac34}
\, ,
\qquad
T > T_{\rm RH}>T_{\rm osc} \, .
\label{rhoshight2}
\eeq
For $T < \Trh$, $T \propto a^{-1}$ and
the energy density $\rho_S$ evolves as
\beq
\rho_S 
\simeq
\frac{1}{2}c_ST^2S_{\rm end}^2
\propto a^{-2}
\, ,
\qquad
T_{\rm RH}>T>T_{\rm osc}
\, .
\label{Eq:OverdampedThermalScaling}
\eeq
Thus when oscillations begin we have
\begin{align}
\rho_S (T_{\rm osc})&=\frac{1}{2}{c_S}  T_{\rm osc}^2 \langle S^2 \rangle_{\rm end} = \frac{3 N \rho_{\rm end}}{128 \pi^2} {c_S}^{2} \beta  \,.
\label{rhostosc2}
\end{align}
This corresponds to the Finite $N$ entry in the 3rd row of Table \ref{Table2}. 
Once $m_{S,\rm th}\gg H$, the condensate enters the adiabatic oscillatory
regime. The adiabatic invariant gives
\beq
\rho_S
\propto
m_{S,\rm th}a^{-3}
\, .
\label{Eq:AdiabaticInvariant2}
\eeq
During radiation domination, $T\propto a^{-1}$ and hence
$m_{S,\rm th}\propto T\propto a^{-1}$, so that
\beq
\rho_S\propto a^{-4}
\, .
\label{Eq:RadiationLikeDilution}
\eeq

This behavior continues until $T=T_\times$ when the energy density is given by
\beq
\rho_S(T_\times)
=
\rho_S(T_{\rm osc})
\left(
\frac{T_\times}
{\Tosc}
\right)^4
\, .
\label{Eq:RhoAtTransitionInitial}
\eeq
Note 
that here 
(and throughout this work) we take $g_*(T_{\rm RH})=g_*(T_{\rm osc})=g_*(T_\times)=106.75$ for our analytic results, though the more precise values for the number of degrees of freedom are used in our numerical results.
Below $T_\times$, the spectator condensate redshifts as non-relativistic
matter. Since the mass is constant, $\rho_S\propto a^{-3}$, and its
present-day energy density is therefore

\bea
\rho_{S,0}
&=&
\rho_S(T_\times)
\left(
\frac{a_\times}{a_0}
\right)^3
\nonumber
\\
&=&
\rho_S(T_\times)
\left(
\frac{T_0}{T_\times}
\right)^3
\frac{g_{*s}(T_0)}
{g_{*s}(T_\times)}
\, .
\label{Eq:RhoPresentFromTransition}
\eea
\noindent
Combining Eqs.~(\ref{rhostosc2}), (\ref{Eq:RhoAtTransitionInitial}) and
(\ref{Eq:RhoPresentFromTransition}) gives
\beq
\rho_{S,0}
=
\rho_S(T_{\rm osc})
\frac{T_0^3T_\times}
{T_{\rm osc}^4}
\frac{
g_{*s}(T_0)
}{
g_{*s}(T_\times)
}
\, .
\label{Eq:RhoPresentCombined}
\eeq
Finally, the relic density is 
\bea
\frac{\Omega_S h^2}{0.12}
\simeq
&&
\left(\frac{m_S}{0.016\,{\rm eV}}\right)
\left(\frac{N}{10^{10}}\right)
\left(\frac{\Hend}{7.4\times10^{12}\,{\rm GeV}}\right)^2
\nonumber
\\
&&
\times
\left(\frac{10^{-8}}{\lambda_{HS}}\right)^{1/2}
\, .
\label{Eq:FinalRelic}
\eea

 However, requiring Eq.~(\ref{Eq:masscondition}) to be satisfied  immediately leads to the conclusion that the expression in Eq.~(\ref{Eq:FinalRelic}) can never be satisfied. For example for $\lambda_{HS} = 10^{-8}$, we must have $m_{S,{\rm eff}} \gtrsim 17$~MeV, leading to an overabundance of spectator dark matter by many orders of magnitude. Since the scalings in Eq.~(\ref{Eq:masscondition}) and in Eq.~(\ref{Eq:FinalRelic}) are the same, the correct relic density can never be satisfied,
thus ruling out the $T_{\rm osc}<T_{\rm RH}$ solution entirely. 
 Furthermore, because this set of solutions is ruled out completely for the finite duration inflation scenario, the analogous solutions for the long inflation scenario where the condensate reaches its equilibrium distribution are also ruled out, since the latter scenario supplies even larger initial conditions for the spectator fluctuations. For completeness, we provide the equilibrium result below. As in the case for $\Trh < \Tosc$, in the limit that $N$ is large we can replace it with its asymptotic joint equilibrium effective value, in which case (for $\kappa = 1$)
\bea
\frac{\Omega_S h^2}{0.12}
&&
\simeq
1.3\times10^{17}
\left(\frac{\Hend}{7.4\times10^{12}\,{\rm GeV}}\right)^2
\left(\frac{m_S}{10^{8}\,{\rm GeV}}\right)
\nonumber
\\
&&
\times
\left(\frac{10^{-8}}{\lambda_{HS}}\right)^\frac32
\, ,
\label{Eq:OmegaQuarticEquilibrium}
\eea
which can be readily obtained by taking $N=(0.05)4\pi^2/\lambda_{HS}$ in Eq.~(\ref{Eq:FinalRelic}).

The above results demonstrate that the thermal reshaping of the condensate 
induced by the Higgs portal is alone insufficient to solve the overproduction problem 
for any value of $N$ which satisfies the isocurvature constraint. 
\emph{The 
inflationary fluctuations}, and hence the initial condensate energy density, 
\emph{are 
simply too large} for thermal reshaping alone to reproduce the observed abundance 
within the viable parameter space.
This rules out scenario 3, in which thermal oscillations are followed by
bare mass oscillations, as a viable solution. However, we will see later that the simple addition of a non-minimal gravitational coupling robustly rescues this solution, such that our labor in this section will not have been in vain. We now turn to scenario 4,
in which the quartic term initially dominates the spectator dynamics
before thermal oscillations take over and are eventually followed by
bare mass oscillations.



\subsection{Quartic $\rightarrow$ thermal $\rightarrow$ bare mass oscillations}
\label{Sec:QTB}

\subsubsection{Finite $N$ inflation}

Our results in the previous subsection assumed that the thermal contribution to the effective mass induced the onset of oscillations of the spectator condensate. However, it is generally possible for the radiatively induced quartic term to first induce oscillations before the condensate subsequently shifts to thermally-induced oscillations. Recall that this corresponds to scenario 4 in the list of possible condensate histories described in Section \ref{OscHistory}. Furthermore, it is important to remember that both the radiative quartic and thermal oscillations are ultimately both controlled by the same portal coupling $\lambda_{HS}$, since we take the bare self-coupling to be zero ($\lambda_{S,0}=0$). Moreover, this sequence of oscillations is only possible when quartic oscillations are induced during the reheating era, not radiation domination. This is because if the quartic term dominates at the onset of radiation domination, it will remain dominant over the thermal mass term at future times. As a result, we restrict ourselves to the case where $T_{\rm osc}>T_{\rm RH}$ in this subsection.

For self-interactions to drive oscillations of the spectator condensate prior to the onset of thermal oscillations, we require that the corresponding effective mass for the quartic term exceeds the thermal mass at the time of oscillation onset. For $\lambda_{S, \rm bare}=0$, such that only the radiatively induced quartic is present, this corresponds to
\beq
S^2 \gtrsim \frac{8 \pi^2}{9 \kappa \lambda_{HS}} (T_{\rm osc}^{(\rm rad)})^2.
\eeq Then, if we assume finite $N$ inflation, the radiation bath temperature at the onset of quartic oscillations during reheating will be given by
\beq
T_{\rm osc}^{(\rm rad)}=\left(\frac{N \kappa \lambda_{HS}^2}{ 16\alpha\pi^4}\right)^{1/8}(H_{\rm end}T_{\rm RH}^2 M_P)^{1/4} \, ,
\label{Tosc4}
\eeq
which yields the lower right entry in Table \ref{tab:oscillation_temperatures_combined} in the limit that $N$ goes to its equilibrium value. 

The condition necessary for the radiative quartic term in the effective mass to induce oscillations prior to the thermal mass is simply $T_{\rm osc}^{(\rm rad)}>T_{\rm osc}^{\rm (th)}$, where the latter can be found in the non-instantaneous reheating entry in Table \ref{Table1}. This directly provides a lower limit on $\lambda_{HS}$ so that oscillations are induced by the one-loop portal quartic coupling for finite $N$ initial conditions of the $S$ fluctuations. We find the limit
\beq
\lambda_{HS}
>
\frac{128\pi^6}{27\sqrt{3}}
\frac{\sqrt{\beta}\,T_{\rm RH}^2M_P}
{\kappa^{3/2}N^{3/2}H_{\rm end}^3}
\, ,
\label{llimlhsq}
\eeq
so that quartic oscillations precede thermal oscillations. 

Next, we can examine the manner in which the energy density of the spectator condensate will redshift after the onset of self-interactions during reheating. In particular, we know from Eq.~(\ref{Eq:AdiabaticInvariant}) along with $\rho_S \propto S^4$ and $m_{S, \rm eff}\propto S$,  that $S^3a^3$ is constant so that $S\propto a^{-1}$ and $\rho_S \propto a^{-4}$. This implies that self-interaction induced 
oscillations of the condensate will be a very efficient mechanism by which the energy density can be redshifted. Interestingly, because $m_{S, \rm quartic} \propto a^{-1}$ and $m_{S, \rm th} \propto a^{-3/8}$, the thermal mass term will often eventually overtake the self-interaction term, such that the two condensate reshaping epochs will occur in series, recalling again that both epochs ultimately originate from the same Higgs portal term in the Lagrangian. 

In other words, we now have the possible sequence: $S$ is frozen at the end of inflation $\rightarrow S$ begins oscillating due to radiatively induced quartic self-interactions $\rightarrow $ thermal mass oscillations $\rightarrow$ bare mass oscillations $\rightarrow$ present day. We will show that this mechanism is also compatible with the correct relic abundance in a large region of parameter space.

Using Eq.~(\ref{Eq:S2endFiniteN}) together with
$\lambda_S=\lambda_S^{\rm rad}$ from Eq.~(\ref{Eq:RadiativeQuarticIntro}),
the spectator energy density at the onset of quartic oscillations with finite $N$ initial conditions takes the form
\beq
\rho_S (a)=\frac{{\kappa} N^2 \lambda_{HS}^2\Hend^4}{1024 \pi^6} \left(\frac{a_{\rm osc}}{a}\right)^4, \hspace{2mm} a_{\rm osc} < a < a_{\rm RH},a_{*} \, ,
\label{Eq:FiniteNQuarticrelic}
\eeq
where we have defined $a_*$ to be the scale factor at which the thermal mass term overtakes the quartic term (if such a scale factor exists for the specified parameters). This behavior of the condensate will continue until one of the other contributions to the effective mass takes over. 

We already considered the case where quartic oscillations are followed directly by bare mass oscillations in Section \ref{Sec:InflationaryMemory}; however, this earlier result was for an isolated self-interacting spectator. The results for the Higgs portal induced quartic $\rightarrow$ bare condensate history are quite similar with the replacement $\lambda_{S,0}\rightarrow\lambda_{S}^{(\rm rad)}$ and a consistency condition requiring thermal oscillations to never be induced. In our results below, we will indeed see that Higgs portal-induced radiative quartic oscillations can be followed either directly by bare mass oscillations or thermal then bare oscillations in the viable parameter space.

We now turn to the case where quartic oscillations are followed by thermal oscillations which are finally followed by bare mass oscillations. It is straightforward to compute the temperature at which the thermal contribution to the effective mass will overtake the quartic contribution (we designate this temperature, $T_*$) by setting $3 \lambda_SS^2=c_ST_{*}^2$. We find
\beq
T_*=\sqrt{\frac{3N\lambda_S}{c_S}}\frac{H_{\rm end}}{2\pi}\left(\frac{a_{\rm osc}}{a_*}\right).
\label{Tstar}
\eeq 
Specializing to the case where $\lambda_S=\lambda_S^{(\rm rad)}$ which is of greatest interest to us and reducing the expression, we find
\beq
T_*=\left(\frac{T_{\rm RH}^4 M_P^2}{H_{\rm end}}\right)^{1/5}\left(\frac{128 \pi^4 \lambda_{HS}}{729 \alpha^2 N\kappa}\right)^{1/10}.
\label{tstar}
\eeq 
 In the case where quartic oscillations are followed by thermal oscillations during reheating, we can determine the final abundance readily using our previous results for the condensate's evolution upon thermally-induced oscillations. We find the following:
\begin{align}
\rho_{S,0}  = & \rho_{S}(T_{\rm osc})\left(\frac{a_{\rm osc}}{a_*}\right)^4\left(\frac{a_*}{a_{\rm RH}}\right)^{27/8}\left(\frac{a_{\rm RH}}{a_\times}\right)^4\left(\frac{a_\times}{a_0}\right)^3 \nonumber \\
 = & \frac{\kappa N^2 \lambda_{HS}^2\Hend^4}{1024 \pi^6}  \left(\frac{T_*}{\Tosc}\right)^\frac{32}{3}\left(\frac{\Trh}{T_*}\right)^{9} \nonumber \\ & \times \left(\frac{T_\times}{\Trh}\right)^4 \left(\frac{T_0}{T_\times}\right)^3\frac{g_{*,s}(T_0)}{g_{*}(T_\times)}
\end{align}  where the expression we have written is for $a_{\rm osc} < a_{\rm RH} < a_\times$, but we recall that the final abundance is the same for the alternative scenario where $a_{\rm osc}<a_\times < a_{\rm RH}$. We have also assumed $g_*(T_{\rm osc})=g_*(T_{*})=g_*(T_{\rm RH})=g_*(T_{\times})$ for simplicity, which will hold in much of our parameter space. Reducing this expression, we find the following result for the final relic abundance
\beq
\Omega_S h^2 = \frac{\pi^2 g_{*,s}(T_0) \sqrt{N}}{240 \sqrt{3 \kappa }\lambda_{HS}}\frac{H_{\rm end} T_{\rm RH} m_S T_0^3}{M_P^2 \rho_c h^{-2}}.
\label{omega4}
\eeq 
Equivalently, with $\kappa =1$
\begin{align}
\frac{\Omega_S h^2}{0.12}&\simeq \left(\frac{N}{10^{10}}\right)^{1/2}\left(\frac{H_{\rm end}}{7.4\times 10^{12} \text{ GeV}}\right) \nonumber \\ & \times \left(\frac{T_{\rm RH}}{1 \text{ TeV}}\right)\left(\frac{m_S}{0.64 \text{ MeV}}\right)\left(\frac{10^{-11}}{\lambda_{HS}}\right)
\label{Eq:FiniteNQuarticThermal}
\end{align} 

This expression for the relic abundance is valid so long as the limit in Eq.~(\ref{llimlhsq}) is satisfied. Moreover, if we take the limit in Eq.~(\ref{llimlhsq}) as an equality and substitute it into Eq.~(\ref{Eq:FiniteNQuarticThermal}), this gives the boundary between 
scenarios 3 and 4 described in Section \ref{Sec:EffectiveMassHierarchy} 
for the case of finite $N$ inflation. However, this boundary lies entirely 
in regions which are excluded by the isocurvature constraint on $N$ and/or 
the limit on the minimum spectator mass. This confirms what we found in 
the previous section, namely that the thermal to bare oscillation branch 
(scenario 3) \emph{strongly over-produces} when $N$ is required to be 
large enough to circumvent the isocurvature bound in 
Eq.~(\ref{Eq:IsocurvatureN}). As a result, the quartic to thermal to bare 
oscillation history is what we are focusing on (scenario 4).

Before we proceed to report and describe our main results for this section, it is worth providing an additional comment on Eq.~(\ref{Eq:FiniteNQuarticThermal}). In particular, it is notable that the form of Eq.~(\ref{Eq:FiniteNQuarticThermal}) is nearly identical to Eq.~(\ref{Eq:k2p4N}) with the replacement $\lambda_S\rightarrow\lambda_{S}^{(\rm rad)}$. This should be unsurprising in light of the adiabatic invariant in Eq.~(\ref{Eq:AdiabaticInvariant}). In particular, the invariant implies that so long as the adiabaticity condition is satisfied between $a_{\rm osc}$ and $a_0$, then the final relic abundance will be sensitive primarily to spectator's energy density and effective mass at the oscillation onset, as well as the effective mass in the present (which is typically the bare mass), and the expansion history. Thus, we find that the quartic $\rightarrow$ thermal $\rightarrow$ bare mass oscillation history gives parametrically the same relic abundance as the quartic $\rightarrow$ bare history (provided that the transitions between the different oscillation epochs retain adiabaticity, which in our case they do). The quartic to thermal transition does numerically alter the abundance by an order 1 coefficient relative to the quartic to bare case, and the oscillation onset itself is transiently non-adiabatic which likewise alters the abundance by an order 1 factor. In order for the thermal mass-induced oscillations to fundamentally alter the final relic abundance parametrically, the thermal mass either needs to drive the onset of oscillations or violate the adiabaticity condition. We will see that in the case of non-minimal gravity in Section \ref{Sec:Nonmin}, the thermal term will in fact drive the onset of oscillations and parametrically alter the final abundance.

\begin{figure*}[!ht]
    \centering
    \includegraphics[
        width=\textwidth
    ]{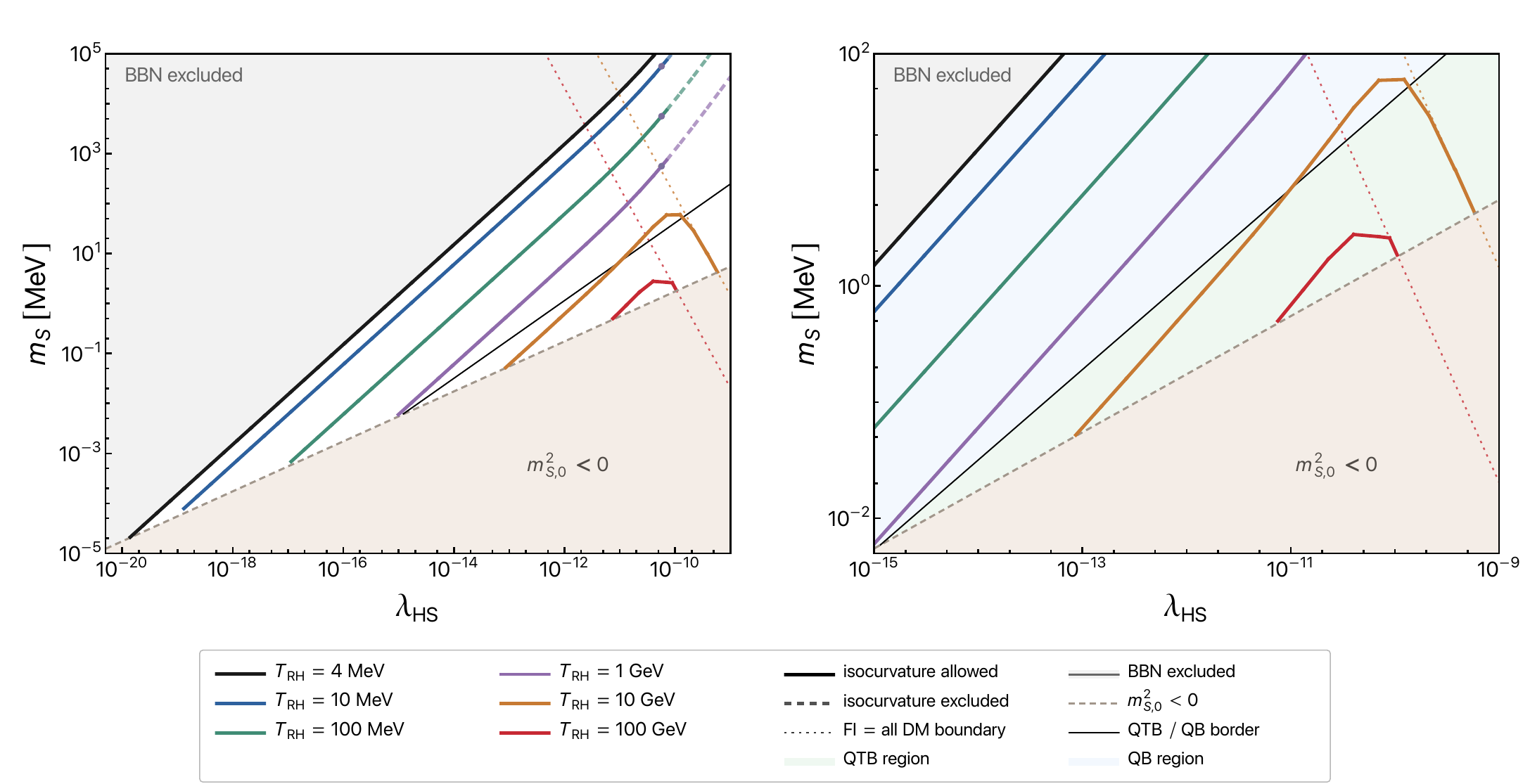}
    \caption{Contours in the $(\lambda_{HS},m_S)$ plane compatible with $\Omega_Sh^2=0.12$ for the Higgs portal model with quartic oscillation onset (for $N=5\times10^{10}$ and several choices of $T_{\rm RH}$). The right panel is a zoom of the left panel with additional shaded regions. The thin black line separates the quartic $\rightarrow$ thermal $\rightarrow$ bare mass oscillation history (shaded green) and the quartic $\rightarrow$ bare mass oscillation history (shaded blue). The beige shaded region corresponds to the constraint from requiring the physical spectator mass to be greater than or equal to the contribution from the Higgs VEV, namely Eq.~(\ref{Eq:masscondition}). The upper left shaded region corresponds to the BBN exclusion from $T_{\rm RH}<4$~MeV. The dashed line segments in the top right of the left panel are excluded from the isocurvature constraint. The dotted red and orange lines in the top right correspond to the pure freeze-in regime for the corresponding value of $T_{\rm RH}$ indicated by the color.}
    \label{Fig:RelicContours}
\end{figure*}

 We now present our results for this subsection. In Fig.~\ref{Fig:RelicContours}, we show contours of
$\Omega_S h^2=0.12$ in the $(\lambda_{HS},m_S)$ plane for fixed
$N=5\times10^{10}$ and several choices of $\Trh$. We neglect the bare
quartic coupling while retaining the radiatively induced
coupling, $\lambda_S^{\rm rad}\propto\lambda_{HS}^2$. As long as the
spectator remains in the finite-$N$ regime, Eq.~(\ref{Eq:FiniteNQuarticThermal})
implies $m_S\propto\lambda_{HS}$, as seen in the figure.

As $\lambda_{HS}$ increases, however, the number of e-folds required to
reach stochastic equilibrium decreases. The joint stochastic
equilibrium result for $N_{\rm eq}$ is given by Eq.~(\ref{jointN}). 
For the fixed value $N=5\times10^{10}$ used in the figure,
$N_{\rm eq}=N$ occurs at $\lambda_{HS}\simeq2\times10^{-11}$.
For larger portal couplings, $N_{\rm eq}<N$, so that the duration of
inflation is sufficient for the spectator fluctuations to saturate their
stochastic equilibrium value. The relic density contour then transitions
from the finite $N$ scaling $m_S\propto\lambda_{HS}$ to the equilibrium
scaling $m_S\propto\lambda_{HS}^{3/2}$ derived below in
Eq.~(\ref{Eq:RelicQuartic}), producing the steeper slope at larger
$\lambda_{HS}$ which is visible in the top right portion of the left panel.

In the left panel, we show an expanded range for $\lambda_{HS}$ and $m_S$.
The black solid line bordering the light gray shaded region in the upper left corresponds to $\Trh = 4$~MeV. In the shaded region, $\Trh < 4$~MeV and is excluded by BBN. 
In the beige shaded portion in the lower part of the figure, the spectator 
mass is below the mass generated by the Higgs VEV and therefore violates 
the bound in Eq.~(\ref{Eq:masscondition}) and is excluded.\footnote{The limit in (\ref{Eq:masscondition}) can be softened slightly, since we really only require that $m_{S,{\rm eff}}^2 > 0$. In particular if this 
is true at the electroweak phase transition, it will always be true. Thus a more accurate limit is $m_{S,{\rm eff}}^2 > \frac12 \lambda_{HS} v^2 (1-T_{\rm EW}^2/v^2)$, where $T_{\rm EW} \simeq 130$~GeV. This 
is about a 10\% correction in the lower bound on $m_S$ shown in the figure. } Above the dotted diagonal lines, the spectator is overproduced by freeze-in as will be discussed below. The position of the dotted lines is $T_{\rm RH}$-dependent. 

We also see a thin black diagonal line defined by setting $T_*$ in Eq.~(\ref{tstar}) to $T_\times$ in Eq.~(\ref{Eq:Tcross}) and solving for $\Trh$ in Eq.~(\ref{omega4}) by setting $\Omega_S h^2 = 0.12$. This gives $m_S \propto \lambda_{HS}^\frac79$ as seen in the figure. Above this line, 
oscillations are initially governed by the quartic and later governed by 
the bare mass as in scenario 2 and $\Omega_S h^2$ is given by 
Eq.~(\ref{Eq:k2p4N}). Below the line, oscillations initially governed by 
the quartic coupling transition to oscillations determined by the thermal 
mass correction before transitioning to bare mass oscillations, and the relic abundance is given by Eq.~(\ref{Eq:FiniteNQuarticThermal}). The dashed portions in the upper right part of the left panel of
Fig.~\ref{Fig:RelicContours} indicate the region excluded by the
small coupling isocurvature constraint. These are to the right of the dots placed on the curves. The right panel of Fig.~\ref{Fig:RelicContours} simply zooms in on the most interesting part of the parameter space.

\begin{figure*}[!ht]
    \centering
    \includegraphics[
        width=\textwidth
    ]{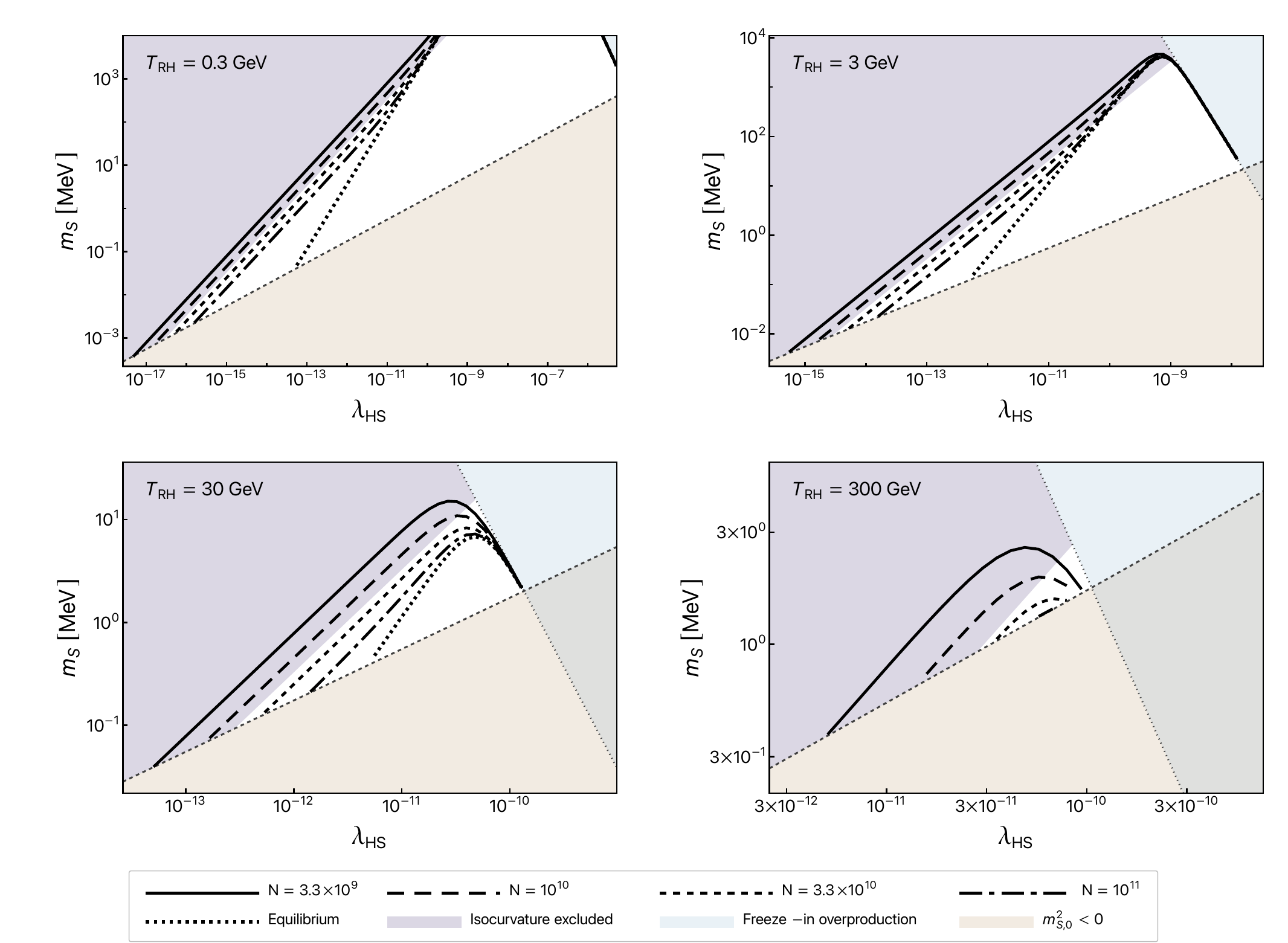}
    \caption{Contours reproducing the observed dark matter abundance,
$\Omega_S h^2=0.12$, in the $(\lambda_{HS},m_S)$ plane for the Higgs portal model for different
durations of inflation $N$, together with the stochastic equilibrium
limit, for four choices of $T_{\rm RH}$. The purple shaded regions are excluded by CMB isocurvature constraints,
while the blue regions at larger portal coupling correspond to
freeze-in overproduction of $S$ particles.
The beige region indicates where the Higgs VEV contribution dominates the zero temperature spectator mass, $m_S^2<\lambda_{HS}v^2/2$. The viable parameter space therefore lies along the relic density contours between the isocurvature, Higgs VEV-dominated, and freeze-in regions.}
    \label{Fig:MasterPlot}
\end{figure*}

In Fig.~\ref{Fig:MasterPlot}, we also plot contours of $\Omega_S h^2 = 0.12$ in the $(\lambda_{HS},m_S)$ plane, but with fixed $\Trh = 0.3, 3, 30$ and 300 GeV in each panel for several values of $N$ above and below the isocurvature limit. In the purple shaded region, the values of $N$ needed to provide the correct relic density fall below the isocurvature limit. The contours shown in the unshaded region satisfy $N>10^{10}$ and follow $m_S \propto \lambda_{HS}$ if $N<N_{\rm eq}$, and $m_S \propto \lambda_{HS}^{3/2}$ for $N=N_{\rm eq}$. The other shaded regions are the same as in Fig.~\ref{Fig:RelicContours}. The unshaded regions therefore define the allowable parameter space for the Higgs portal-induced condensate reshaping mechanism highlighted in this work, including scenario 4 mentioned in Section \ref{Sec:EffectiveMassHierarchy}. The dotted curve corresponds to the value of $N$ at the equilibrium limit. 
Since that value of $N$ depends on $\lambda_{HS}$, for certain values of $\Trh$, we reach the equilibrium limit and the curves merge as seen in the top two panels of the figure. We also see as $\Trh$ increases, the allowed region of parameter space closes and most of the parameter space is allowed only for relatively low reheating temperatures, below the electroweak scale. Generally, such low reheating temperatures make baryogenesis significantly more challenging, but see \cite{Garcia:2026ulw} for a detailed study of leptogenesis in the setting of low reheating temperatures.

\subsubsection{Long inflation: stochastic equilibrium}

We now turn to the case where the spectator reaches its stochastic equilibrium variance of fluctuations during inflation. In addition to this case providing different initial conditions for the spectator fluctuations at the end of inflation, the onset of quartic oscillations and the subsequent transition to thermal mass domination will also occur at distinct scale factors (and associated bath temperatures) from the finite $N$ case. Taking into account the 
initial conditions for the $S$ fluctuations from the joint inflationary 
evolution, we can determine the scale factor at which quartic oscillations 
during reheating begin.  Starting with Eq.~(\ref{Eq:QuadraticOscillation}) 
and $H = \Hend (\aend/a_{\rm osc})^\frac32$, and using ${m_{S,\rm eff}} = \sqrt{3 \lambda_S} S$ with $S$ given by Eq.~(\ref{Eq:JointEvolutionS}) 
we arrive at 
\beq
\frac{a_{\rm osc}}{a_{\rm end}}\simeq 17 ( \lambda_{HS})^{-1/3},
\eeq 
where we took $\kappa = 1$. After a short non-adiabatic transition period around the onset of oscillations,
the subsequent evolution is approximately adiabatic. This allows us to again use the 
adiabatic invariant in Eq.~(\ref{Eq:AdiabaticInvariant}) to readily compute an analytic estimate of the final relic abundance. In particular, the adiabatic invariant gives 
\beq
\rho_{S,0}\simeq \rho_S (a_{\rm osc})\left(\frac{m_{S,\rm eff}(a_0)}{m_{S,\rm eff}(a_{\rm osc})}\right)\left(\frac{a_{\rm osc}}{a_0}\right)^3 \, .
\eeq 
In the case of quartic oscillation onset during reheating we can convert the ratio of scale factors to temperatures, remembering that prior to reheating $T \propto a^{-\frac38}$ and subsequently $T \propto a^{-1}$. Then using Eq.~(\ref{Tosc4}) for $\Tosc$ and using 
equilibrium initial conditions supplied by the joint stochastic inflationary evolution of $S$ and $h$ we find
\beq
\Omega_S h^2 \simeq 0.093 \kappa^{-1/2} \lambda_{HS}^{-3/2}\frac{T_{\rm RH} m_S T_0^3 H_{\rm end}}{M_P^2 \rho_c h^{-2}}.
\label{Eq:RelicQuartic}
\eeq 
Putting this all together, our final expression for the relic abundance is
\begin{align}
\frac{\Omega_S h^2}{0.12}&\simeq \left(\frac{10^{-11}}{\lambda_{HS}}\right)^{3/2}\left(\frac{T_{\rm RH}}{100 \text{ GeV}}\right) \nonumber \\ & \times \left(\frac{H_{\rm end}}{7.4\times10^{12} \text{ GeV}}\right)\left(\frac{m_S}{2 \text{ MeV}}\right).
\end{align}
This expression provides the $m_S \propto \lambda_{HS}^\frac32$ behavior alluded to earlier. 

The solutions for $\Omega_S h^2 = 0.12$ for the equilibrium solutions are seen in Fig.~\ref{Fig:MasterPlot} displayed as the dotted lines. As noted earlier these lines scale as $m_S \propto \lambda^\frac32$ as found in Eq.~(\ref{Eq:RelicQuartic}).
For completeness, the temperature $T_*$ in this case can be obtained from Eq.~(\ref{Tstar}) with the replacement $N \to (.16)^2 4 \pi^2/\lambda_{HS}$.

Finally, at sufficiently large $\lambda_{HS}$, production from the thermal
bath becomes important and eventually dominates over the primordial
condensate contribution. In the Higgs-decay-dominated freeze-in regime,
$m_S<m_h/2$, the relic abundance scales approximately as
\beq
\Omega_S^{\rm FI}h^2\propto\lambda_{HS}^2m_S
\, ,
\eeq
\noindent
and hence the relic-density contour follows
\beq
m_S\propto\lambda_{HS}^{-2}
\, .
\eeq
\noindent
This relation, as well as the value of $\lambda_{HS}$ above which freeze-in
overtakes the primordial-condensate contribution, is discussed in
Appendix~\ref{app:FI}.
This explains the sharp turnover and the common negative slope of the
contours at large $\lambda_{HS}$. Physically, increasing the portal
coupling enhances freeze-in production and therefore increases the
number density of spectator particles. Maintaining the observed dark
matter abundance consequently requires a smaller spectator mass. The
turnover thus marks the transition from a relic abundance controlled
primarily by the inflationary condensate to one controlled by thermal
freeze-in production.

Before ending this subsection, we provide a brief comment on the relationship between our numerical results shown in the figures and our numerous analytic expressions. We note that our analytic expressions for the relic abundance are approximate results which assume an immediate transition from a frozen $S$ field to adiabatic oscillations. In reality, there will be a brief non-adiabatic transition period which modifies the result by an $\mathcal{O}(1)$ factor. In addition, for the cases where the quartic term induces oscillations, our analytic expressions neglect another order one factor resulting from the transition from quartic to quadratic oscillations. We determined the associated numerical factors in each case, such that our analytic expressions could be refined by multiplying the RHS of any of our results by the corresponding factor. For the first factor resulting from the transient non-adiabatic period near oscillation onset, there are actually four characteristic numerical factors depending on whether quartic or thermal oscillations are induced first, and whether $T_{\rm osc}>T_{\rm RH}$ or $T_{\rm osc}<T_{\rm RH}$. For the case where quartic oscillations are induced first, we found the numerical factor to be 2.2 for $T_{\rm osc} > T_{\rm RH}$ and 1.4 for $T_{\rm RH}>T_{\rm osc}$. For the case where thermal oscillations are induced first, we find 0.67 for $T_{\rm osc}>T_{\rm RH}$ and 0.25 for $T_{\rm osc}<T_{\rm RH}$. For the factor resulting from the quartic to quadratic transition, we find 2.7. Each of these factors was found to be quite robust across the majority of our parameter space. Note that it is also possible to obtain the latter factor analytically by conserving the comoving oscillator action. If one wishes to account for these numerical factors, one can simply multiply any of our relic expressions by one or both of these factors (depending on which regime is considered) to obtain refined estimates which match even more closely to the corresponding results in the figures.

\subsection{Non-minimal gravity}
\label{Sec:Nonmin}

Lastly, we consider a simple modification of our Higgs portal model by adding a non-minimal gravitational coupling of the form
\beq
\mathcal{L}\supset - \frac{1}{2}\xi_S R S^2,
\eeq which gives an additional contribution to the effective spectator mass of the form \cite{Garcia:2025rut}
\beq
m_{S,\rm eff}^2 \supset \xi_S R = \begin{cases} 12 \xi_S H^2, \text{ inflation} \\ 3\xi_S H^2, \text{  reheating} \end{cases},
\eeq where on the RHS we have used the general result $R=3(1-3w)H^2$ for a flat FLRW universe with a constant background equation of state. We will see that this simple modification dramatically changes the evolution of the spectator condensate and the resulting parameter space. In particular, we will see that the thermal $\rightarrow$ bare mass oscillation sequence is rescued, high reheating temperatures become allowed (and even preferred), the $T_{\rm RH}>T_{\rm osc}$ hierarchy becomes viable, and we gain orders of magnitude in both $\lambda_{HS}$ and $m_S$. We note that non-minimal gravitational interactions have previously been studied for spectator dark matter in other contexts \cite{Fairbairn:2018bsw,Garcia:2023qab}. 

The effective mass term generated by the curvature coupling is significantly larger than the other two contributions to the effective spectator mass (bare mass and quartic) during inflation for the light spectator masses and feeble $\lambda_{HS}$ of interest in this work. As a result, the non-minimal gravitational coupling will dictate the stochastic inflationary equilibrium variance of the spectator's fluctuations, which takes the form
\beq
\langle S^2 \rangle_{\rm eq} \simeq \frac{H_{\rm end}^2}{32 \pi^2 \xi_S}.
\eeq After inflation has ended, $S$ will evolve mildly during the reheating period due to the non-minimal gravitational coupling such that generically $S_{\rm RH}\neq S_{\rm end}$ even if oscillations are not induced until radiation domination. Once reheating has completed, the non-minimal gravitational contribution to the effective mass goes to zero. The $S$ field will then remain frozen at $\sim S_{\rm RH}$ until the onset of oscillations by the thermal mass. Since the initial size of the spectator fluctuations is suppressed relative to the case where the inflationary dynamics are determined by the spectator self-interactions, it will be the thermal mass which is responsible for the onset of spectator oscillations rather than the radiative quartic. Our approach for computing an analytic estimate for the final abundance will therefore first be to determine the ratio $S_{\rm RH}/S_{\rm end}$. Then we will use $S_{\rm RH}$ as an input to estimate the initial energy density at the onset of oscillations before using the adiabatic invariant to determine the final abundance.

To determine $S_{\rm RH}/S_{\rm end}$, we first recall that after inflation has ended, the evolution of $S$ during reheating due to the non-minimal gravitational coupling will be dictated by the following equation of motion
\beq
\ddot{S}+3H\dot{S}+\xi_S R S=0,
\eeq when the curvature term dominates the effective mass. After a change of variables to  $x=\ln(a/a_{\rm end})$, we have
\beq
S''+\frac{3}{2}S'+3\xi_S S = 0, 
\eeq where $'$ denotes a derivative with respect to $x$. This admits the solution
\beq
S(x)=S_{\rm end}(A_+e^{r_+ x} +A_- e^{r_- x}),
\eeq with characteristic exponents
\beq
r_{\pm}=\frac{-3/2\pm\sqrt{9/4-12\xi_S}}{2}.
\eeq The fast mode with exponent $r_-$ will quickly decay away such that the evolution is determined by $r_+\simeq -2\xi_S$. We will initialize the reheating era evolution using $S(0)=S_{\rm end}$ and the drift velocity inherited from the inflationary slow mode, namely
$S'(0)\simeq -(3/2-\nu)S_{\rm end}$ with $\nu=\sqrt{9/4-12\xi_S}$. Taking the initial velocity to be zero would only change the final abundance by $\mathcal{O}(10\%)$. One then finds $A_+\simeq(1-4\xi_S/3)$ such that
\beq
S(x)\simeq S_{\rm end}(1-4\xi_S/3)e^{-2\xi_S x}.
\eeq after the fast mode has decayed. We then find the ratio we sought to determine, namely
\beq
\frac{S_{\rm RH}}{S_{\rm end}}\simeq \left(1-\frac{4\xi_S}{3}\right)\left(\frac{\alpha T_{\rm RH}^4}{3M_P^2 H_{\rm end}^2}\right)^{\frac{2}{3}\xi_S}.
\label{Eq:Srh}
\eeq

With Eq.~(\ref{Eq:Srh}), we are equipped with the necessary initial condition to evolve the condensate from the start of radiation dominated era through the present, recalling that the non-minimal gravitational coupling term will drop out during radiation domination since $R=0$. After the onset of thermal oscillations, the energy density of the spectator condensate will redshift like radiation, as shown in Section \ref{Sec:thermalToBare}, and in particular Eq.~(\ref{Eq:RadiationLikeDilution}). We can then use the adiabatic invariant in Eq.~(\ref{Eq:AdiabaticInvariant}) to obtain an expression for the final relic abundance. We first consider the case where $T_{\rm RH}>T_{\rm osc}$ such that
\begin{align}
\rho_{S,0}&\simeq\rho_S({a_{\rm osc}})\left(\frac{m_S}{m_{S,\rm eff}(a_{\rm osc})}\right)\left(\frac{a_{\rm osc}}{a_0}\right)^3 \nonumber \\ &=\frac{g_{*,s}(a_0)}{240\sqrt{6}\xi_S \sqrt{\lambda_{HS}}}\frac{H_{\rm end}^2 m_S T_0^3}{M_P^2} \left(\frac{S_{\rm RH}}{S_{\rm end}}\right)^2.
\end{align}
We find the abundance to be
\begin{align}
\frac{\Omega_S h^2}{0.12}&\simeq\left(\frac{m_S}{1.4 \text{ MeV}}\right)\left(\frac{H_{\rm end}}{7.4\times10^{12} \text{ GeV}}\right)^{2-8\xi_S/3} \nonumber \\ &\times\left(\frac{10^{-12}}{\lambda_{HS}}\right)^{1/2}\left(\frac{T_{\rm RH}}{10^{12}~{\rm GeV}}\right)^{16\xi_S/3} \, ,
\label{Eq:NonminRelic1}
\end{align}
where we used $\xi_S = 0.03$ for specificity while leaving the explicit $\xi_S$ dependence in the exponents of $T_{\rm RH}$ and $H_{\rm end}$. If we include the numerical factor pertaining to the non-adiabatic transition period near the onset of oscillations, the DM mass in the above expression can be re-calibrated to $m_S\sim 6$~MeV which is in close agreement with the figure.

Through a similar derivation process, we can find the corresponding result for the $T_{\rm osc}>T_{\rm RH}$ case, with the major differences being that the temperature evolution during reheating is distinct, and that rather than $S_{\rm RH}/S_{\rm end}$, we require $S_{\rm osc}/S_{\rm end}$. We find
\begin{align}
\frac{\Omega_S h^2}{0.12}&\simeq\left(\frac{m_S}{1.5 \text{ MeV}}\right)\left(\frac{H_{\rm end}}{7.4\times10^{12} \text{ GeV}}\right)^{2-\frac{8}{3}\xi_S} \nonumber \\ & \times \left(\frac{T_{\rm RH}}{10^8 \text{ GeV}}\right)^{\frac{1}{3}+\frac{16}{9}\xi_S}\left(\frac{10^{-15}}{\lambda_{HS}}\right)^{\frac{2}{3}-\frac{16}{9}\xi_S},
\label{Eq:NonminRelic2}
\end{align} where again we have taken $\xi_S=0.03$ but left factors of $\xi_S$ explicit in the power laws to illustrate the dependencies.

Next, we consider the isocurvature power spectrum which becomes 
\begin{align}
\mathcal{P}(k_*)&\simeq \frac{8 m_{S,\rm eff}^2}{3H^2} \exp\left[\frac{-4N_* m_{S,\rm eff}^2}{3H^2}\right] \nonumber \\ & =32 \xi_S e^{-16 \xi_S N_*}.
\label{Eq:IsoNonminGrav}
\end{align} For $\xi_S=0.03$ and $N_*=55$ we find
\beq
\mathcal{P}(k_*)\simeq 3.3\times 10^{-12}, 
\eeq which is safely below the isocurvature bound from Planck. If we use Eq.~(\ref{Eq:IsoNonminGrav}) to find an isocurvature bound on $\xi_S$ from Planck by requiring $\mathcal{P}(k_*)<8.3\times10^{-11}$, we find $\xi_S \gtrsim 0.026$ for $N_*=55$. Because we take $\xi_S>0.026$ in our displayed results below, there will be no isocurvature constraints shown on the plots, contrary to our prior results. Note that considering the finite $N$ scenario in this case is less interesting since the stochastic equilibrium variance is already reached at relatively low values of $N$, in contrast to the scenarios we studied in previous sections where the effective spectator mass was much lighter. 

\begin{figure*}[!ht]
    \centering
    \includegraphics[
        width=\textwidth
    ]{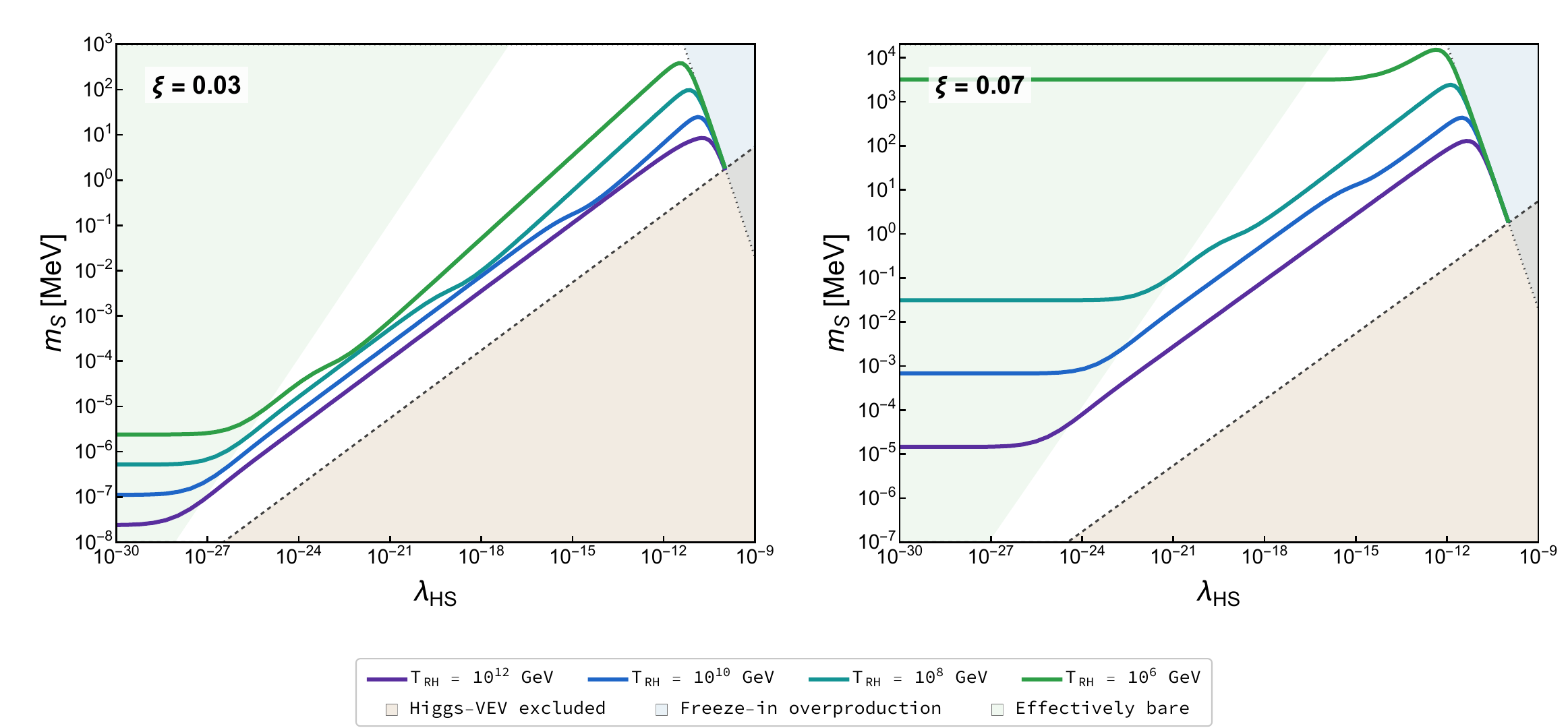}
    \caption{Contours in the $(\lambda_{HS},m_S)$ plane compatible with $\Omega_Sh^2=0.12$ for a feeble Higgs portal interaction with a non-minimal gravitational coupling term for four choices of $T_{\rm RH}$, and each panel uses a distinct choice of $\xi_S$. The beige shaded regions correspond to $m_S^2<\lambda_{HS}v^2/2$, while the light blue shaded regions correspond to Higgs decay freeze-in over-production. The white regions correspond to a condensate history characterized by Higgs-portal induced thermal oscillations followed by bare mass oscillations. The light green shaded regions are allowed, but no Higgs portal-induced thermal or quartic oscillations occur in this region (effectively corresponding to a standard quadratic spectator).}
    \label{Fig:NonminGrav}
\end{figure*}

We show the resulting contours consistent with $\Omega_S h^2=0.12$ in the $(\lambda_{HS},m_S)$ plane in Fig. \ref{Fig:NonminGrav} for two choices of $\xi_S$ and four choices of $T_{\rm RH}$. The interesting parameter space for the Higgs-portal induced thermal reshaping of the condensate is shown in the white regions with each contour corresponding to a distinct choice of $T_{\rm RH}$. The blue and beige shaded regions correspond to the freeze-in and Higgs VEV-induced mass bounds respectively, which were also shown in previous plots. The light green shaded region corresponds to the parameter space where thermal oscillations are never induced, but rather only bare mass oscillations, such that the DM behaves as an isolated spectator. The analytic result in Eq.~(\ref{Eq:NonminRelic1}) corresponding to $T_{\rm osc}<T_{\rm RH}$ is best seen for the $T_{\rm RH}=10^{12}$~GeV purple contour in the left panel. For this high choice of $T_{\rm RH}$, we are always in the $T_{\rm RH}>T_{\rm osc}$ regime such that $m_S \propto \lambda_{HS}^{1/2}$ throughout the white region. This is one of the key parametric results associated with this subsection. Note that this parametric behavior is fundamentally different from that of a self-interacting spectator, such that we have demonstrated that the Higgs portal-induced thermal oscillations are a novel mechanism for reshaping spectator condensates. We emphasize that there is no regime observed in this scenario where the radiative quartic term induces onset of oscillations, such that self-interactions play no role here. For the $T_{\rm osc}>T_{\rm RH}$ case, we can examine the contours with lower values of $T_{\rm RH}$ in the left panel. For instance, the $T_{\rm RH}=10^6$~GeV and $10^8$~GeV contours transition from obeying the $m_S\propto\lambda_{HS}^{1/2}$ scaling at low values of $\lambda_{HS}$ (which corresponds to $T_{\rm RH}>T_{\rm osc}$ shown in Eq.~(\ref{Eq:NonminRelic1})) to the steeper $m_S\propto \lambda_{HS}^{0.61}$ scaling (for $\xi_S=0.03$) which corresponds to $T_{\rm osc}>T_{\rm RH}$ shown in Eq.~(\ref{Eq:NonminRelic2}).

In the right panel which depicts $\xi_S=0.07$, we see that choosing a larger value of $\xi_S$ allows for higher dark matter masses for a fixed choice of $T_{\rm RH}$. This is because higher $\xi_S$ decreases the initial fluctuation size at the end of inflation and decreases the ratio $S_{\rm RH}/S_{\rm end}$. These two effects allow for smaller condensate fluctuations at the onset of radiation domination which in turn allows for larger dark matter mass.  

We note that in the results shown in our contour plot we have used an oscillation-averaged Ricci scalar during the reheating period. We have numerically checked that using the full oscillatory Ricci scalar only changes the final abundance by approximately 4\%.

In summary, the addition of a non-minimal gravitational coupling impacts our story in two ways. The first and most important effect is that for typical values of $\xi_S$, a non-minimal gravity term will substantially increase the effective spectator mass during inflation, which leads to a significant reduction of the amplitude of the fluctuations at the end of inflation. When this reduced amplitude for the fluctuations is paired with a feeble Higgs portal, we find that the thermal mass of the spectator will induce early onset of oscillations, causing the spectator condensate's energy density to redshift as radiation before transitioning to bare mass oscillations. Because the amplitude $S_i$ is reduced, the effective mass associated with the radiatively induced quartic term  $m_{S,\rm eff}^2 \propto S^2$ pays a sharp penalty relative to the thermal mass term, such that it becomes negligible compared to the thermal mass over the entire parameter space, and never drives oscillations. Recall that this is in sharp contrast to the previous scenario without the non-minimal coupling where the initial fluctuations were quite large and thus we found that the quartic term initiated oscillation onset in all of our viable parameter space. The reduced initial amplitude also reduces the requirement for long reheating periods, such that the reheating temperature can be much higher when a non-minimal gravity term is present. The parameter space is also expanded considerably, as seen in Fig.~{\ref{Fig:NonminGrav}}, where the allowed range of both $m_S$ and $\lambda_{HS}$ are expanded by about 5 and 10 orders of magnitude respectively relative to the case without non-minimal gravity. The second, and less important, effect is that the curvature coupling induces a minor evolution of the condensate during the reheating period. 

\section{Conclusions}
\label{Sec:Conclusion}

Light scalar fields generically acquire large stochastic fluctuations during
inflation. The long wavelength modes of these fluctuations satisfy the equation of motion of a homogeneous scalar condensate. If stable, these condensates naturally survive until late times
and, in the simplest quadratic scenario, this leads to a catastrophic
overproduction of dark matter. For a weak-scale spectator and an inflationary scale with
$\rho_{\rm end}\simeq10^{63}\,{\rm GeV}^4$, the resulting relic abundance exceeds
the observed value by more than twenty orders of magnitude (cf. Eq.~(\ref{Eq:OmegaQuadraticb})).

Here, we have shown that an extremely feeble Higgs-portal interaction provides a simple and natural solution to this problem. The Higgs portal induces both a radiatively generated self-interaction and a thermal mass for the spectator, both of which can lead to early oscillations and efficient redshifting of the spectator condensate's energy density to achieve the correct relic abundance. This defines a new cosmological role for the Higgs portal. Unlike
freeze-out, the dark matter never thermalizes, and unlike
freeze-in, the portal interaction does not produce the dark matter abundance through decays or scattering. Instead, the Higgs portal is responsible for draining the energy density stored in a large, pre-existing inflationary condensate, determining how much of it survives to the present day. We expect that this general principle could be exploited with other interactions between the dark and visible sectors. 

The scalar dark matter candidate we consider is a Standard Model gauge singlet and communicates with the visible sector exclusively through the Higgs portal term $\lambda_{HS}(H^\dagger H)S^2/2$. The spectator then has three significant contributions to its effective mass which can each drive post-inflationary oscillations, namely $m_{S,\rm eff}^2 \supset m_S^2+3\lambda_S^{(\rm rad)}S^2+c_ST^2$, where both $\lambda_{S}^{(\rm rad)}$ and $c_S$ are determined by $\lambda_{HS}$ (see Eqs.~(\ref{Eq:RadiativeQuarticIntro}) and (\ref{Eq:cs})). We find that when the stochastic fluctuations generated during inflation are very large, it is the radiatively induced quartic self-interaction (with effective mass $m_{S,\rm eff}^2\supset 3\lambda_S^{(\rm rad)}S^2$) which will generally lead to the onset of the first oscillations. After quartic oscillations, it is then possible for the condensate to transition either directly to bare mass oscillations, or instead to thermal oscillations followed by bare mass oscillations (these correspond to scenarios 2 and 4, respectively, described in Section \ref{OscHistory}). Both histories can reproduce the observed dark matter abundance, as illustrated in Figs.~\ref{Fig:RelicContours} and \ref{Fig:MasterPlot}.

When the thermal term overtakes the radiative quartic
contribution, it drives the condensate into an effectively quadratic, but time-dependent, potential. Although the thermal oscillations are quadratic, their evolution differs qualitatively from that induced by the constant bare mass $m_S$. In particular, because $m_{S,\rm eff}^{(\rm th)}\propto T$ decreases with time,  the condensate dilutes faster than ordinary non-relativistic matter. During radiation domination, $\rho_S\propto a^{-4}$, while during matter-like reheating (where
$T\propto a^{-3/8}$) we find $\rho_S\propto a^{-27/8}$. This
enhanced dilution continues until the low-temperature bare mass takes over, after which the
usual matter-like scaling $\rho_S\propto a^{-3}$ is recovered. The thermal
contribution temporarily increases the instantaneous spectator energy density through a term of order $\rho_{S,\rm th}\simeq c_S T^2S^2/2$. However, the earlier onset of oscillations and subsequent dilution can compensate for this increase. Once the oscillatory evolution becomes adiabatic, the corresponding invariant in Eq.~(\ref{Eq:AdiabaticInvariant}) allows the successive quartic, thermal, and bare-mass regimes to be connected smoothly and provides a simple description of the final relic abundance.

If, at the end of inflation, the amplitude of spectator fluctuations is relatively small, the thermal mass term can be responsible for the initial onset of oscillations rather than the quartic. We study this case in detail in the setting of a non-minimal gravitational coupling in Section \ref{Sec:Nonmin}. The curvature coupling leads to a suppressed inflationary amplitude of fluctuations. The  associated parameter space extends to significantly smaller DM masses and portal couplings compared to the case without non-minimal gravity (for instance, compare the plot ranges in Figs. \ref{Fig:MasterPlot} and \ref{Fig:NonminGrav}).

In this work, we considered both the possibility that inflation was sufficiently long for the light scalar to reach its equilibrium distribution of fluctuations (the standard case), as well as the less commonly studied alternative where inflation lasts a much smaller number of e-folds $N\ll N_{\rm eq}$. Indeed, one of the major problems facing a non-self-interacting scalar is due to the assumption of a large initial variance, $\langle S^2\rangle$, given by Eq.~(\ref{Sisat}). This, however, assumes an equilibrium distribution for the fluctuations, which often requires a very long duration of inflation-- for instance, $\sim 10^{22}$ e-folds for a weak scale mass with $H\sim 10^{13}$~GeV. For models of inflation which are eternal \cite{Vilenkin:1983xq,Linde:1986fd}, this may not be a problem. However, for even a slight deformation of a model of inflation such as the Starobinsky model \cite{Staro}, the number of e-foldings ($N \sim H \tau$)
becomes finite \cite{Antoniadis:2025pfa}. 
For an isolated quadratic spectator, finite $N$ solutions yielding the correct relic abundance can be found for low reheating temperatures as seen in Eq.~(\ref{Eq:OmegaFiniteN}) and previously observed in \cite{Feiteira:2025rpe}. However, we show that such solutions are in conflict with the isocurvature limit on $N$. When a bare quartic self-coupling is present
we find that a fairly wide range of masses, couplings, and reheating temperatures are possible as seen in Eq.~(\ref{Eq:k2p4N}) and Fig. \ref{Fig:RelicContoursQuartic}. 

We also find that the finite $N$ inflation scenario opens additional parameter space for our Higgs-portal mechanism (see Fig.~\ref{Fig:MasterPlot}). In the long inflation regime, where the spectator reaches its stochastic
equilibrium distribution, the relic density contour follows the
characteristic scaling $m_S\propto\lambda_{HS}^{3/2}$.
By contrast, for a finite duration of inflation the reduced initial
spectator variance leads to $m_S \propto \lambda_{HS}$. The two branches therefore reflect physically distinct inflationary initial conditions, while both benefit from the subsequent dilution induced by the Higgs portal. Importantly, however, we find that the Planck bound on isocurvature places a major constraint on the shorter inflation scenario for either an isolated or coupled spectator, as shown in Eq.~(\ref{Eq:IsocurvatureN}), such that the anticipated gain from reducing the amplitude of fluctuations is limited.
 
At sufficiently large $\lambda_{HS}$, thermal production eventually
overtakes the primordial condensate contribution. The relic density
contours then turn over in the ($\lambda_{HS}, m_S$) plane and approach the usual freeze-in-dominated behavior (see Figs. \ref{Fig:RelicContours}, \ref{Fig:MasterPlot}, and \ref{Fig:NonminGrav}).
There is, however, an important difference relative to the standard freeze-in approach where the initial dark matter abundance is usually assumed to be
negligible. For a spectator scalar, inflation generically provides an
irreducible primordial population, and this contribution must therefore be
consistently included. We have shown that there exists a region of parameter
space in which freeze-in can nevertheless determine the final dark matter
abundance, while the Higgs portal simultaneously dilutes, and eventually
\emph{erases the memory of the inflationary production}. Thus, freeze-in
remains a viable mechanism even in the presence of the
inflationary initial condition, provided that the portal interaction is
sufficiently efficient at removing its primordial contribution, along with any additional dilution from lower reheating temperatures. This is one
of the important results of the present work.

In summary, we have shown that a feeble Higgs portal coupling (feeble enough that the spectator condensate remains intact and that freeze-in is negligible) can significantly affect the evolution and final relic spectator abundance. 
The portal coupling, $\lambda_{HS}$,
generates both a quartic interaction through loop corrections, as well as a thermal mass from the presence of Higgs bosons in the radiation bath. The initial conditions for the condensate at the end of inflation are generally altered by the coupling. Instead of an initial amplitude given by Eq.~(\ref{Sisat}), the initial conditions must be determined from the joint distribution as explained in Appendix \ref{app:joint} and given by Eq.~(\ref{Eq:JointEvolutionS}). 
After inflation, the radiative quartic and/or thermal masses suppress the inflationary abundance twice: they can trigger oscillations earlier than the bare mass would, and, once oscillations begin, they cause the condensate to redshift faster than non-relativistic
matter. This enhanced dilution continues until the bare mass eventually dominates.  This mechanism opens a qualitatively new region of parameter space for spectator dark matter. In the parameter space explored here, viable solutions extend from DM masses of order 10 GeV down to sub-eV levels, well below the typical GeV-scale emphasized in earlier spectator dark matter analyses.

Future developments include a more complete treatment of the
finite-temperature effective potential in the presence of a large coherent
background, as well as the transition to the dissipative regime in which
interactions with the thermal bath eventually destroy the coherent
condensate altogether.

\subsection*{Acknowledgments}
We would like to thank Marcos Garcia for helpful conversations.
This project has received support from the European Union's Horizon 2020 research and innovation program under the Marie Sklodowska-Curie grant agreement No 860881-HIDDeN.
  The work of K.A.O.~was supported in part by DOE grant DE-SC0011842  at the University of
Minnesota. S.E.H. was supported by the Doctoral Dissertation Fellowship awarded by the University of Minnesota. Y.M. acknowledges support by Institut Pascal at Université Paris-Saclay during the Paris-Saclay Astroparticle Symposium 2025, with the support of the P2IO Laboratory of Excellence (program “Investissements d’avenir” ANR-11-IDEX-0003-01 Paris-Saclay and ANR- 10-LABX-0038), the P2I axis of the Graduate School of Physics of Université Paris-Saclay, as well as the CNRS IRP UCMN.

\section*{Appendices}
\appendix

\setcounter{section}{0} 
\setcounter{subsection}{0}
\renewcommand{\thesubsection}{\Alph{subsection}}

\counterwithin{equation}{subsection} 

\makeatletter
\@addtoreset{equation}{subsection}

\newcommand{\autolabel}[1]{%
  \protected@edef\@currentlabel{\thesubsection}%
  \label{#1}%
}
\makeatother

\subsection{Thermal evaporation of the condensate}
\label{app:evap}

In our treatment, we have assumed that the spectator scalar develops stochastic fluctuations during inflation, followed by quartic and/or thermal oscillations driven by the Higgs portal interaction, followed lastly by a period of bare mass oscillations. This assumes that the scalar behaves as a coherent condensate throughout its evolution. It is therefore necessary to determine that the condensate does not undergo fragmentation or thermal evaporation. Because the radiatively induced quartic term in our model is so feeble, the direct fragmentation rate of the condensate due to anharmonic oscillations is very small (see Appendix C of Ref. \cite{Garcia:2025rut}).

However, thermal evaporation of the condensate is at least conceptually more plausible \cite{Nurmi:2015ema}. We can obtain an estimate of the thermal evaporation rate by considering the elastic scattering process
\beq
S_0 + H \rightarrow S_p + H,
\eeq where $S_0$ and $S_p$ are the zero mode of the $S$ condensate and the $S$ particle respectively. The amplitude squared and cross section for the contact interaction are simply
\beq
|\mathcal{M}_{SH\rightarrow SH}|^2=\lambda_{HS}^2, \text{ and } \sigma_{SH\rightarrow SH}\simeq\frac{\lambda_{HS}^2}{16 \pi s}.
\eeq The scattering rate per zero mode quantum is then
\beq
\Gamma_{SH \rightarrow SH} = 4 \int_0^\infty \frac{d^3 p}{(2\pi)^3}f(E_H)\frac{p}{E_H}\frac{\lambda_{HS}^2}{16 \pi s},
\eeq where $f(E_H)$ is the Bose distribution and $s=m_H(T)^2+m_S(T)^2+2E_H m_S(T)$. In the high temperature regime where thermal evaporation may be relevant, the masses of both the Higgs and the scalar are dominated by their thermal masses rather than their bare masses. As a result, the Mandelstam variable $s$ becomes $s=c_HT^2+c_S T^2+2\sqrt{p^2+c_H T^2}\sqrt{c_S}T$, where $c_H$ is the coefficient of the thermal contribution to the Higgs mass, $m_{H,{\rm eff}}^2 \supset c_H T^2$, and depends on Standard Model gauge and Yukawa couplings as well as $\lambda_H$ and $\lambda_{HS}$. In the feeble Higgs portal coupling regime ($\lambda_{HS}\lesssim 10^{-7}$) where the spectator never thermalizes, we have $s\simeq c_HT^2$; however, we do not take this approximation in the following calculation. For this thermal evaporation rate to become significant, we would need to have $\Gamma_{SH\rightarrow SH}/H\gtrsim 1$. We then notice that $\Gamma_{SH\rightarrow SH}/H$ will scale as $1/T$ (or $1/T^3$ during reheating), such that the ratio becomes larger at lower temperatures and late times. Therefore, to determine whether thermal evaporation of the condensate is significant, it is suitable to check whether the ratio $\Gamma_{SH\rightarrow SH}/H$ is small at the lowest relevant temperature, which is near the electroweak scale. We find the following
\beq
\frac{\Gamma_{SH\rightarrow SH}}{H}|_{T=160\text{ GeV}}\approx 9\times 10^{-3}\left(\frac{\lambda_{HS}}{10^{-8}}\right)^2,
\eeq where we used $c_H\simeq0.4$. As a result, we find that thermal evaporation of the condensate from the elastic scattering process mentioned above is negligible for our parameter space of interest. Perhaps unsurprisingly, it is apparent from the above analysis that the thermal evaporation rate only becomes significant for approximately the same coupling strength at which the $S$ particles thermalize. We note that the above calculation does not capture all possible thermal dissipation effects (see, for example, \cite{Banerjee:2025qyg}). We leave a more detailed treatment of dissipative effects for future work.

\subsection{Joint stochastic evolution of the $S$ and $H$ field fluctuations during inflation}
\label{app:joint}

Naively, in our spectator setup one might be tempted to write $\langle S^2 \rangle_{\rm end}$ using its standard equilibrium distribution at the end of inflation. However, due to the Higgs portal coupling we introduced in Eq.~(\ref{Eq:PortalPotential}), a proper treatment should involve a multi-field stochastic evolution of the fluctuations of both $S$ and $H$ during inflation. The purpose of this section, therefore, is to perform the two-field treatment with the goal of determining the proper initial conditions to use for the post-inflationary evolution.

Recall that treating $S$ as an isolated field gives the equilibrium form of $\langle S^2\rangle_{\rm eq }$ as follows \cite{Starobinsky:1994bd}
\beq
\langle S^2 \rangle_{\rm eq,iso} = \frac{\int_{-\infty}^{+\infty}S^2 \exp [-\frac{8\pi^2 V(S)}{3 H^4}]dS}{\int_{-\infty}^{+\infty} \exp [-\frac{8\pi^2 V(S)}{3 H^4}]dS}.
\eeq For $V(S)=\frac{1}{2}m_S^2S^2$, this gives the familiar result shown in Eq.~(\ref{Sisat}) with $m_{S,\rm eff}\Rightarrow m_S$. In the case where the quartic term is instead dominant, the equilibrium variance takes the form in Eq.~(\ref{Eq:Sendquartic}). However, when the Higgs portal coupling in Eq.~(\ref{Eq:PortalPotential}) is introduced, we instead must consider the expectation value of this operator using the full joint potential $V(H,S)$, which gives us 
\beq
\langle S^2 \rangle_{\rm eq,joint} = \frac{\int_{-\infty}^{+\infty}\int_{0}^{+\infty} h^3 S^2 \exp \left[-\frac{8\pi^2 V(H,S)}{3 H^4}\right] dh\, dS}{\int_{-\infty}^{+\infty}\int_{0}^{+\infty} h^3 \exp \left[-\frac{8\pi^2 V(H,S)}{3 H^4}\right] dh\, dS},
\label{Eq:JointExpectationS2}
\eeq 
where we have used the $O(4)$ radial Jacobian for the Higgs and defined $h$ from $H^\dagger H=h^2/2$. We emphasize that Eq.~(\ref{Eq:JointExpectationS2}) is the stationary de Sitter equilibrium variance, requiring approximately constant Hubble parameter and enough inflation to reach equilibrium.

Then, we take $V(H,S)$ to be the following:
\begin{align}
V(H,S) & =\frac{1}{2}m_S^2 S^2 + \frac{1}{2}\lambda_{HS}S^2 H^\dagger H +{\lambda_H}(H^\dagger H)^2 \nonumber \\ & +\frac{\kappa \lambda_{HS}^2}{64 \pi^2}S^4,
\label{Eq:JointHSPotential}
\end{align} 
where we have used Eq.~(\ref{Eq:RadiativeQuarticIntro}). Note that we do not consider a bare self-interaction coupling $\lambda_{S}$, but only the unavoidable radiative quartic term induced by the Higgs portal. We have also neglected the electroweak Higgs mass term since in the inflationary regime, the Higgs quartic term will be dominant by many orders of magnitude. 

We will compute the above equilibrium variance for the joint potential in the limit where the bare mass is subdominant relative to the other two contributions to the spectator effective mass, which is the relevant limit for the scenario of interest in this work. Using Eqs.~(\ref{Eq:JointExpectationS2}) and (\ref{Eq:JointHSPotential}) we find
\beq
\langle S^2 \rangle_{\rm eq,joint} \simeq 0.05 \frac{H^2}{\lambda_{HS}}, \hspace{4mm} \langle |S|\rangle_{\rm eq,joint} \simeq 0.16 \frac{H}{\sqrt{\lambda_{HS}}},
\label{Eq:JointEvolutionS}
\eeq 
for $\lambda_H=0.01$, or more generally
\beq
 \langle |S|\rangle_{\rm eq,joint} \simeq 0.51 \lambda_H^{1/4}\frac{H}{\sqrt{\lambda_{HS}}},
\label{Eq:JointEvolutionS2}
\eeq
where we have taken $\kappa=1$. For $\lambda_H=0.01$ the result for $\langle S^2 \rangle_{\rm eq, joint}$ is suppressed by approximately a factor of 30 relative to the isolated quartic spectator. 
This then alters the equilibrium number of e-folds to that given in Eq.~(\ref{jointN}).
Importantly, the appropriate averaged quantity when determining analytic estimates of the relic abundance is either $\langle S^2\rangle_{\rm end}$ or $\langle|S|\rangle_{\rm end}$ depending on whether the abundance $\Omega_Sh^2 =\rho_{S,0}/ \rho_ch^2$ is proportional to $S_{\rm end}$ or $S_{\rm end}^2$. In our case, when the radiative quartic term induces the onset of oscillations, the final abundance scales as $S_{\rm end}$ so we use $\langle |S|\rangle_{\rm end}$ in that case, whereas for thermal or bare mass oscillation onset, we use $\langle S^2 \rangle_{\rm end}$. Similarly, for the radial Higgs field, we have
\beq
\langle h^2 \rangle_{\rm eq,joint} = \frac{\int_{-\infty}^{+\infty} dS \int_{0}^{+\infty} dh \hspace{1mm} h^5 \exp \left[-\frac{8\pi^2 V(H,S)}{3 H^4}\right]}{\int_{-\infty}^{+\infty} dS \int_{0}^{+\infty} dh \hspace{1mm} h^3 \exp \left[-\frac{8\pi^2 V(H,S)}{3 H^4}\right]}.
\label{Eq:JointExpectationH2}
\eeq For $\lambda_H=0.01$, the above expression numerically evaluates to $\langle h^2 \rangle_{\rm eq,joint}\simeq (2.9-3.5)H^2$ when tested over a broad range of $m_S$ and $\lambda_{HS}$. The analogous result for an isolated Higgs potential is $\langle h^2 \rangle_{\rm eq,iso}\simeq 3.5H^2$, demonstrating that the joint potential does not significantly impact the variance of Higgs fluctuations at the end of inflation, even though we have seen that the joint distribution of $S$ fluctuations may be significantly suppressed by the joint treatment.

We can further support the claim that the portal interaction $\propto \lambda_{HS}h^2S^2$ primarily governs the equilibrium distribution in our regime of interest by computing the result for a simplified potential which excludes the spectator's bare mass and self-interaction terms entirely. Specifically, if we use $\tilde{V}(H,S)=\lambda_{HS}h^2S^2/4+\lambda_Hh^4/4$, we can compute the integrals in Eq.~(\ref{Eq:JointExpectationS2}) analytically to obtain
\beq
\langle S^2 \rangle_{\widetilde{\rm eq}}=\frac{\sqrt{6}}{4\pi}\frac{\Gamma(1/4)}{\Gamma(3/4)}\sqrt{\lambda_H}\frac{H^2}{\lambda_{HS}}
\label{Eq:JointEvolutionS3}
\eeq
\beq
\langle h^2 \rangle_{\widetilde{\rm eq}}=\sqrt{\frac{3}{2\pi^2}}\frac{\Gamma(5/4)}{\Gamma(3/4)}\frac{H^2}{\sqrt{\lambda_H}}.
\label{Eq:JointEvolutionS4}
\eeq If we take $\lambda_H=0.01$, we find $\langle S^2 \rangle_{\widetilde{\rm eq}}=0.058H^2/\lambda_{HS}$ and $\langle h^2 \rangle_{\widetilde{\rm eq}}=2.9H^2$. Thus, when using the simplified potential $\tilde{V}(H,S)$, we recover a result which is nearly the same as the full result. 

\subsection{Upper bound on $\lambda_{HS}$ from Higgs portal freeze-in}
\autolabel{app:FI} 

The Higgs portal-induced condensate reshaping mechanism described above will not be responsible for the final spectator abundance if the Higgs portal coupling $\lambda_{HS}$ is sufficiently large. For large couplings, eventually the standard Higgs portal freeze-in process will be sufficiently strong to produce the correct abundance. For even larger couplings, the DM will equilibrate with the SM radiation bath and freeze-out will be operative, but we do not consider such large couplings here. As a result, our mechanism of interest will be operative when the Higgs portal coupling is smaller than a particular $m_S$-dependent value, which we now derive.

After EWSB, we have
\beq
H=\frac{1}{\sqrt{2}}\begin{pmatrix} 0 \\ v+h
\end{pmatrix}
\eeq where $h$ is the physical Higgs. 
In particular, the Higgs portal interaction term in Eq.~(\ref{Eq:PortalPotential}) generates a term of the form 
\beq
V(S,H) \supset \frac{\lambda_{HS}v}{2} h S^2,
\eeq which is responsible for $h \rightarrow SS$. We note that because the light spectator masses of interest in this work satisfy $m_S < m_h/2$, the decay channel $h \rightarrow SS$ will be the dominant freeze-in channel, rather than $2\rightarrow 2$ annihilations. The corresponding decay rate is 
\beq
\Gamma_{h \rightarrow SS}=\frac{\lambda_{HS}^2v^2}{32 \pi m_h}\sqrt{1-\frac{4 m_S^2}{m_h^2}}.
\eeq 

Because the majority of the DM production for Higgs portal freeze-in decay occurs in the $T\lesssim m_h$ regime, we use Maxwell-Boltzmann statistics. Using $x=m_h/T$, we have
\beq
\frac{dY_S}{dx}=\frac{135\sqrt{10}}{2\pi^5 g_{*}^{3/2}}\frac{M_P \Gamma_{h\rightarrow SS}}{m_h^2}x^3 K_1(x).
\eeq Integrating to $x=\infty$ and enforcing the correct abundance gives us the condition we intended to determine 
\beq
\frac{\Omega_S^{\rm FI} h^2}{0.12}\simeq \left(\frac{\lambda_{HS}}{10^{-10}}\right)^2 \left(\frac{m_S}{20 \text{ MeV}}\right),
\label{Eq:FreezeInRelic}
\eeq 
where this result holds for $T_{\rm RH}\gtrsim m_h$. There is a separate result for $T_{\rm RH} \ll m_h$, which applies to some of our results above for low reheating temperatures. In this case, we have
\begin{align}
Y_S(T_{\rm RH})&=\frac{180\sqrt{10}}{\pi^5 g_{*}(T_{\rm RH})^{3/2}}\frac{\Gamma_{h\rightarrow SS} M_P}{m_h^2} \nonumber \\ & \times \left(\frac{T_{\rm RH}}{m_h}\right)^7\int_0^{x_{\rm RH}}dx \hspace{1mm} x^{10} K_1 (x),
\end{align} where $x_{\rm RH}=m_h/T_{\rm RH}$. We then find the following expression for the final relic abundance
\beq
\frac{\Omega_S^{\rm FI} h^2}{0.12}\simeq \left(\frac{\lambda_{HS}}{10^{-10}}\right)^2 \left(\frac{m_S}{20 \text{ MeV}}\right)\left(\frac{T_{\rm RH}}{18 \text{ GeV}}\right)^7
\label{Eq:FreezeInRelic2}
\eeq 
for $T_{\rm RH} \ll  m_h$. In particular, while $T_{\rm RH}=18$~GeV is a natural normalization for comparison with Eq.~(\ref{Eq:FreezeInRelic}), one should not take $T_{\rm RH}\gtrsim 8$~GeV in Eq.~(\ref{Eq:FreezeInRelic2}) if one wishes for better than 10\% accuracy. The above expressions in Eqs.~(\ref{Eq:FreezeInRelic}) and (\ref{Eq:FreezeInRelic2}) can be used to place a contour in the $(\lambda_{HS},m_S)$ plane which constrains the viable parameter space to be below the contour to avoid overproducing dark matter by freeze-in. We have depicted these freeze-in constraints in our key figures illustrating our model's viable parameter space.



\end{document}